\documentclass[11pt,a4paper]{article}
\pdfoutput=1

\usepackage{jheppub}

\usepackage{amsfonts}
\usepackage{mathtools}
\usepackage{amssymb}
\usepackage{eucal}
\usepackage{mathrsfs}
\usepackage{slashed}
\usepackage{amsmath}
\usepackage{braket}
\usepackage{graphicx}
\usepackage{float}
\usepackage{subfloat}
\usepackage{relsize}
\usepackage[font=scriptsize,labelfont=bf]{caption}
\usepackage{subcaption}
\usepackage{xfrac}
\usepackage{enumerate}
\usepackage{wasysym}

\usepackage{ifthen}
\usepackage[dvipsnames]{xcolor}
\usepackage{tikz}
\usepackage{mathtools}
\usepackage{xspace}

\newcommand{\TRH}{T_{\rm RH}}

\newcommand{\ie}{{\em i.e.,}\xspace}

\newcommand{\GeV}{{\rm GeV}\xspace}

\newcommand{\keV}{{\rm keV}\xspace}

\DeclareMathOperator{\LH}{L}
\DeclareMathOperator{\PRH}{P_R}
\DeclareMathOperator{\PLH}{P_L}

\newcommand{\injFact}{\Delta _\text{EMD}\xspace}

\newcommand{\mDM}{ m_{\rm DM}\xspace}

\newcommand{\lrb}[1]{\left( #1 \right)}

\newcounter{NumArgs}

\newcommand{\eqs}[1]{\setcounter{NumArgs}{0}\foreach\i in{#1}{\stepcounter{NumArgs}}%
	\ifthenelse{\equal{\theNumArgs}{1}}{eq.~(\ref{#1})}%
	{\ifthenelse{\equal{\theNumArgs}{2}}%
		{eqs.~\foreach\i[count=\q]in{#1}{\ifthenelse{\equal{\q}{\theNumArgs}}{and (\ref{\i})}{(\ref{\i})~}}}%
		{eqs.~\foreach\i[count=\q]in{#1}{\ifthenelse{\equal{\q}{\theNumArgs}}{and (\ref{\i})}{(\ref{\i}),~}}}}}

\newcommand{\Eqs}[1]{\setcounter{NumArgs}{0}\foreach\i in{#1}{\stepcounter{NumArgs}}%
	\ifthenelse{\equal{\theNumArgs}{1}}{Eq.~(\ref{#1})}%
	{\ifthenelse{\equal{\theNumArgs}{2}}%
		{Eqs.~\foreach\i[count=\q]in{#1}{\ifthenelse{\equal{\q}{\theNumArgs}}{and (\ref{\i})}{(\ref{\i})~}}}%
		{Eqs.~\foreach\i[count=\q]in{#1}{\ifthenelse{\equal{\q}{\theNumArgs}}{and (\ref{\i})}{(\ref{\i}),~}}}}}

\newcommand{\refs}[1]{\setcounter{NumArgs}{0}\foreach\i in{#1}{\stepcounter{NumArgs}}%
	\ifthenelse{\equal{\theNumArgs}{1}}{(\ref{#1})}%
	{\ifthenelse{\equal{\theNumArgs}{2}}%
		{\foreach\i[count=\q]in{#1}{\ifthenelse{\equal{\q}{\theNumArgs}}{and (\ref{\i})}{(\ref{\i})~}}}%
		{\foreach\i[count=\q]in{#1}{\ifthenelse{\equal{\q}{\theNumArgs}}{and (\ref{\i})}{(\ref{\i}),~}}}}}

\newcommand{\Figs}[1]{\setcounter{NumArgs}{0}\foreach\i in{#1}{\stepcounter{NumArgs}}%
	\ifthenelse{\equal{\theNumArgs}{1}}{Fig.~\ref{#1}}%
	{\ifthenelse{\equal{\theNumArgs}{2}}%
		{Figs.~\foreach\i[count=\q]in{#1}{\ifthenelse{\equal{\q}{\theNumArgs}}{and \ref{\i}}{\ref{\i}~}}}%
		{Figs.~\foreach\i[count=\q]in{#1}{\ifthenelse{\equal{\q}{\theNumArgs}}{and \ref{\i}}{\ref{\i},~}}}}}

\title{Freeze-in baryogenesis and early matter domination}

\author[1]{Ioannis  Dalianis}
\author[2]{, Andreas Goudelis}
\author[3]{\!\!, Dimitrios Karamitros}
\author[1]{\!\!, Pantelis Papachristou}
\author[1]{\!\!, Vassilis C. Spanos}
\affiliation[1]{Section of Nuclear and Particle Physics, Department of Physics, \\
National and Kapodistrian University of Athens, 
   GR-157 84 Athens, Greece}
\affiliation[2]{Laboratoire de Physique de Clermont (UMR 6533), CNRS/IN2P3, Univ.\ Clermont Auvergne, 4 Av.\ Blaise Pascal, F-63178 Aubi\`ere Cedex, France}
\affiliation[3]{School of Physics and Astronomy, The University of Manchester, Manchester M13 9PL, United Kingdom}
\emailAdd{idalianis@phys.uoa.gr}
\emailAdd{andreas.goudelis@clermont.in2p3.fr}
\emailAdd{dimitrios.karamitros@manchester.ac.uk}
\emailAdd{pantelisp@phys.uoa.gr}
\emailAdd{vspanos@phys.uoa.gr}

\abstract{The freeze-in mechanism has been shown to allow the simultaneous generation of cosmic dark matter and a viable matter-antimatter asymmetry in the universe. When the underlying interactions are described by higher-dimensional, non-renormalizable operators, the relevant freeze-in processes take place close to the highest considered cosmic temperatures. In this paper we study how the presence of a fluid that temporarily dominates the energy content of the 
early
universe affects the predictions of this ``Ultraviolet Freeze-In Baryogenesis'' scenario. We find that this additional cosmic component has a significant impact on the predictions of concrete microscopic models, allowing for reheating temperatures which are much lower than those required in the simplest 
cosmological scenario. Moreover, we show that inflationary observables can constrain the parameter space of such models, once the latter are examined in conjunction with concrete models of inflation.}

\begin{document}
\maketitle

\section{Introduction}\label{sec:intro}
\setcounter{equation}{0}
The nature of dark matter (DM) and the origin of the baryon asymmetry of the universe are two of the most important questions at the interface between particle physics and cosmology. In \cite{Hall:2010jx} it was suggested that a common framework for the simultaneous explanation of both could be found within the context of freeze-in DM \cite{McDonald:2001vt,Hall:2009bx}: in freeze-in scenarios, DM interacts so weakly (``feebly'') with the Standard Model (SM) and its thermal bath that the relevant interactions never reach thermal equilibrium, an element which is in line with the third Sakharov condition \cite{Sakharov:1967dj} required for successful baryogenesis. This idea was exploited in \cite{Hall:2010jx} in the context of asymmetric frozen-in DM and, later, in \cite{Shuve:2020evk,Berman:2022oht} in the context of symmetric DM candidates which can oscillate in a mechanism reminiscent of (albeit not identical with) ARS leptogenesis \cite{Akhmedov:1998qx}. 
    
In \cite{Goudelis:2021qla,Goudelis:2022bls} an alternative approach was proposed in which the out-of-equilibrium $CP$-violating decays or scatterings of heavy, exotic bath particles are responsible for the generation both of the observed DM abundance and of an asymmetry in the SM lepton sector. The latter can afterwards be translated into a baryon asymmetry through the electroweak sphaleron transitions. In the case of decays the underlying interactions were taken to be renormalizable \cite{Goudelis:2021qla}, leading to the freeze-in of both DM and the asymmetry at low temperatures. In  \cite{Goudelis:2022bls}, on the other hand, the possibility of non-renormalizable interactions was studied, which brings the entire setup in the realm of ``Ultraviolet (UV) freeze-in'' (in the context of DM see \textit{e.g} \cite{Elahi:2014fsa} and on the baryogenesis side \cite{Baldes:2014rda,PhysRevLett.113.181601,Baldes:2015lka}), \textit{i.e.} freeze-in which occurs at the highest considered cosmic temperature -- typically identified with the reheating temperature of the universe. In both cases, the scenario tends to predict light DM, with a mass of the order of a few keV. Moreover, in the case of \cite{Goudelis:2022bls}, the values of the reheating temperature that were required in order to achieve successful DM and baryon asymmetry production were strikingly high, especially if the underlying dynamics are described by a dimension-5 operator.

The reason for this appears to be the fact that a viable baryon asymmetry tends to be generated alongside an overproduced DM. Indeed, while the asymmetry is generated through the interference of tree-level and one-loop contributions in perturbation theory, the DM is predominantly produced at tree-level. One is then lead to the choice of increasing the reheating temperature in order to achieve successful baryogenesis while drastically decreasing the DM mass to avoid overclosure of the universe.

The previous findings were obtained within the context of the standard thermal history for the universe, \textit{i.e.} assuming that after the complete decay of the inflaton field(s)  radiation domination settled in and continued uninterrupted until matter domination. It is, however, known that the early universe thermal history is rather unknown and there might be departures from this simple picture. A well-motivated and studied scenario, that changes the underlying thermodynamics  drastically in the pre-BBN era, is that of a transit non-thermal period  where  a pressureless and short-lived fluid  dominates the energy density of the universe \cite{Kawasaki:1999na}. Indeed, assuming that a fluid-dominant period is interjected during radiation domination, the decay of the fluid will dilute both the DM and a potential baryon asymmetry. 

In this work we will place ourselves in such a scenario: we will assume that, during some period after the end of inflation, the energy density of the universe was dominated by a fluid which subsequently decayed bringing along a dilution of all pre-existing abundances in the plasma. We will see that indeed, in this situation the predictions of the ``Ultraviolet freeze-in baryogenesis'' scenario presented in \cite{Goudelis:2022bls} can be modified substantially and that the mechanism can be efficient for a quite heavier dark matter mass and assuming a much lower reheating temperature. This is possible because the DM abundance and the size of baryon asymmetry scale differently with temperature. 
    
At the same time, late entropy production induced by the decay of the fluid can have a non-negligible impact on inflationary observables. The observed number of e-folds of inflation $N_*$ is sensitive both to inflationary dynamics as well to postinflationary evolution. Indeed, the power spectrum of primordial density perturbations is not strictly scale-invariant, and any modification of the expansion rate affects the rate at which each mode $k$ reenters the horizon and, thus, the CMB-measured $n_s(k)$ value. In this paper we will explicitly express how the e-folds number $N_*$ can be related with the underlying parameters of two simple microscopic models. In the framework of specific inflationary models this correlation can be translated  into expected values of inflationary observables. In particular, the predicted values for the spectral index of the two-point correlation function of the primordial scalar perturbations, $n_s$, and the tensor-to-scalar ratio, $r$, are shifted by a particular amount if late entropy injection takes place \cite{Liddle:2003as, Kinney:2005in, Martin:2010kz, Easther:2013nga, Dalianis:2018afb}. 

The  advantage of this scenario is that the values of the reheating temperature and the dilution size are automatically correlated due to microphysical requirements. 
We will show that for the allowed range of reheating temperatures and dilution sizes that lead to a viable DM abundance and baryon asymmetry, the predictions of some typical inflationary models are confined in a smaller range of values on the $(n_s, r)$ contour plane. We will also show that dimension-5 and dimension-6 operators lead to different predictions, and therefore interesting conclusions can be inferred concerning the compatibility between concrete microscopic and cosmological models. 
In order to be quantitatively specific  we will make use of some minimal and general models of inflation, namely large field models, small field models and general models with a $R^2$-like plateau, so as to illustrate  the interplay between particle physics models and cosmology. We mention that our line of arguments and analysis are general and can be also applied to other models that predict a baryon asymmetry and DM production. 

The paper is organised as follows: In Section \ref{sec:UVFIBG} we review the main features of symmetric DM freeze-in baryogenesis in its UV-dominated version and briefly introduce the models that will be studied in the following. In Section \ref{sec:thefluid} we show how the predicted abundance is modified in the presence of a decaying fluid. In Section \ref{sec:observables} we describe how the inflationary observables are modified by the entropy that is injected upon the fluid's decay, focusing on some representative models of inflation. Lastly, in Section \ref{sec:conclusions} we summarize our main findings and conclude.

\section{Ultraviolet-dominated freeze-in baryogenesis}\label{sec:UVFIBG}

Let us start by briefly reviewing the main features of the UV-dominated freeze-in baryogenesis mechanism, as presented in \cite{Goudelis:2022bls}. The starting point assumption is that the DM particle species only interacts feebly with the bath particles through an effective operator $\mathcal{\hat{O}}_{\left(n\right)}$ of mass dimension $(n+4)$
\begin{equation}
\mathcal{L}\,\supset\,\frac{1}{\Lambda^n}\,\mathcal{\hat{O}}_{\left(n\right)} \;,
\end{equation}
where $n=1,2,\ldots$ and $\Lambda$ is the energy scale of the effective field theory (EFT). This operator is responsible for the simultaneous generation of the freeze-in DM abundance and of the baryon asymmetry of the universe. Due to the non-renormalizable nature of the underlying interactions, both are mostly produced in the ultraviolet regime, \textit{i.e.} close to the reheating temperature $T_{\rm RH}$. 

In the models that we will study, the DM is produced via pair-annihilations of bath particles. In the limit that $\Lambda$ is much larger than the masses involved, the DM yield obtained by solving the corresponding Boltzmann equation can be written in a fairly compact form \cite{Elahi:2014fsa}
\begin{align}
    Y_{\rm DM}(T) \approx  2 A \, \frac{4^{n+1} n! (n+1)!}{ 2n-1 } \dfrac{45}{ 1024\times 1.66 \, \pi^7 g_{* s}\sqrt{g_{*\rho}}}
    \frac{M_{\rm Pl} \,  (T_{\rm RH}^{2n-1} -  T^{2n-1})}{\Lambda^{2n}} \, ,
\label{eq:YDMT}
\end{align}
where $A$ is a model-dependent constant, $M_{\rm Pl}$ is  the reduced Planck mass and $g_{* s},\,g_{*\rho}$ are the relativistic degrees of freedom of the plasma associated with the entropy and energy densities, respectively.

As pointed out in the introduction, the driving factor for studying the simultaneous generation of freeze-in DM and baryon asymmetry is that the relevant interactions are always out-of-equilibrium, so that the third Sakharov is inherently satisfied. Additionally, the $CP$ symmetry is explicitly broken through the complex couplings appearing in the effective operators. The most non-trivial part of the proposed baryogenesis mechanism is the violation of the baryon number. In our case the effective interactions are total lepton (and baryon) number-conserving resulting to sector-wise lepton asymmetries generated in the SM, $L_{\text{SM}}$, and an exotic sector, $L_{\text{ex}}$, while the total lepton asymmetry remains zero, $L=L_{\text{SM}}+L_{\text{ex}}=0$. However, if the exotic particles are taken to be $SU(2)_{\LH}$-singlets the electroweak sphalerons, which violate $B+L$ but conserve $B-L$, only affect the non-zero lepton asymmetry stored in the SM sector. Hence, they ``see" a non-vanishing lepton asymmetry and convert it into a baryon one, which is proportional to $Y_B\propto Y_{B-L_{\text{SM}}}$. Once they depart from equilibrium at $T=T_{\text{sph}}$ the baryon and lepton asymmetries are separately conserved and the baryon asymmetry remains frozen at the value $Y_B\propto Y_{B-L_{\text{SM}}}|_{T_{\text{sph}}}$, which, in principle, is non-zero.  

In order to present the corresponding predictions for the baryon asymmetry we will introduce two specific models: one based on a scalar DM candidate produced through a dimension-5 operator, and that of a fermionic candidate produced through a dimension-6 one.

\subsection{Dimension-5 operator}

The first case study that we consider is an extension of the SM by a complex gauge-singlet scalar field $\varphi$, which is our DM candidate, along with two heavy vector-like Dirac fermions $F_{1,2}$, which are singlets under $SU(3)_c \times SU(2)_{\LH}$ but carry hypercharge. The interaction Lagrangian reads
\begin{equation}\label{eq:Lagdim5}
\mathcal{L}_{\text{int}}=\mathcal{L}_{\text{gauge}} + \frac{\lambda_{1}}{2 \Lambda}\left(\bar{e}\PLH F_{ 1}\right)\varphi^*\varphi^*+\frac{\lambda_{2}}{2 \Lambda}\left(\bar{e}\PLH F_{ 2}\right)\varphi^*\varphi^*+\frac{\kappa}{\Lambda^2}\left(\bar{e}\PLH F_{ 1}\right)\left(\bar{F}_{2}\PRH e\right)+\text{h.c.}\,,
\end{equation}
where $e$ can be any of the Standard Model charged leptons and $\mathcal{L}_{\text{gauge}}$ contains the gauge interactions of the heavy fermions $F_i$. The latter keep the $F_i$ in equilibrium with the SM bath particles. Besides, the heavy fermions carry the same lepton number as the SM leptons, and so total lepton number is conserved. Additionally, a $\mathbb{Z}_3$ symmetry is imposed, under which both the $F_i$'s and $\varphi$ are charged, to ensure DM stability.

    \begin{figure}[t!]
    \centering
    \begin{subfigure}[t]{.55\linewidth}
      \includegraphics[width=\linewidth]{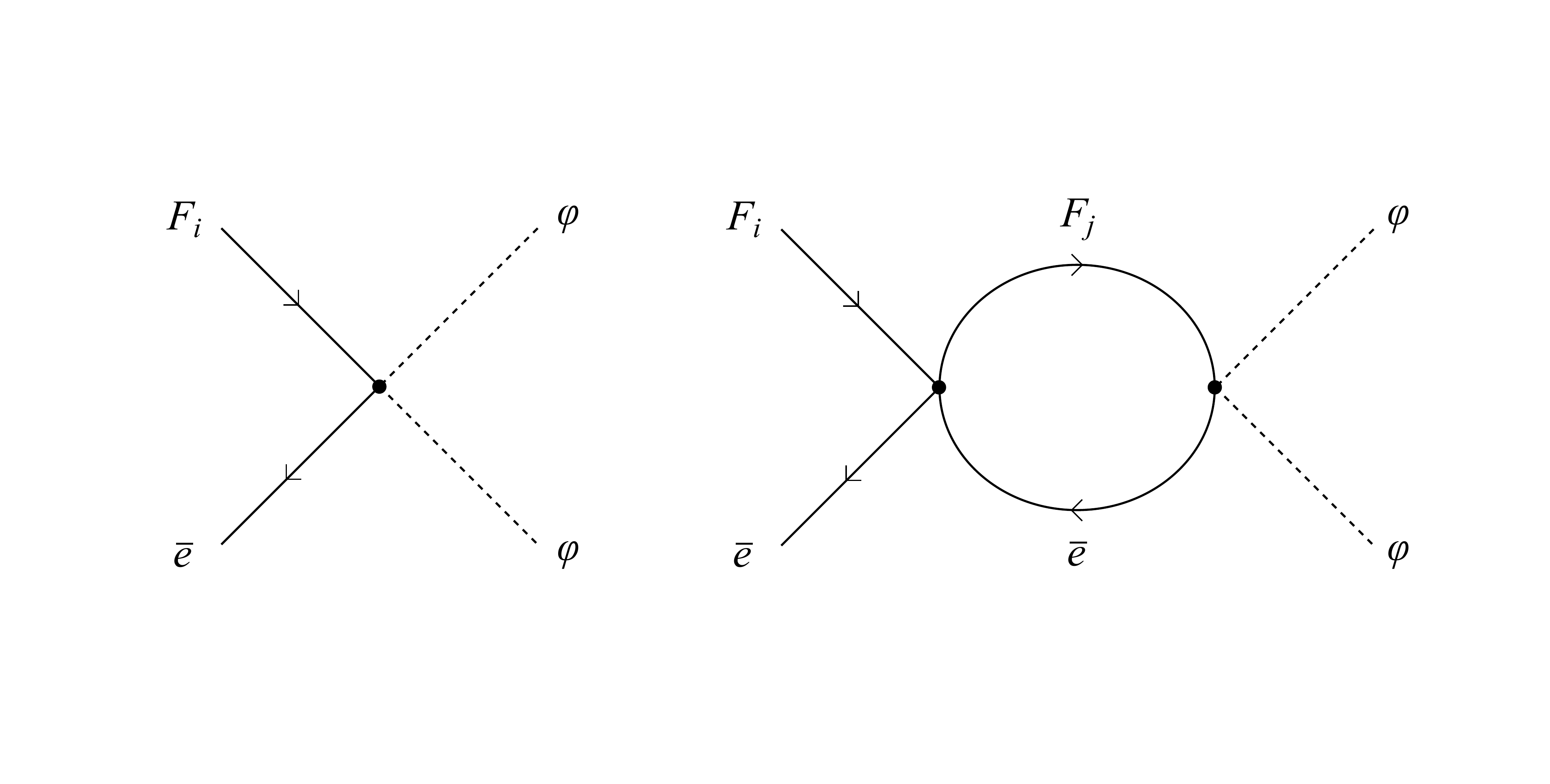}
      \caption{}
      \label{fig:Feyndiagrams}
    \end{subfigure}\hspace*{\fill}
    \begin{subfigure}[t]{.45\linewidth}
    \includegraphics[width=\linewidth]{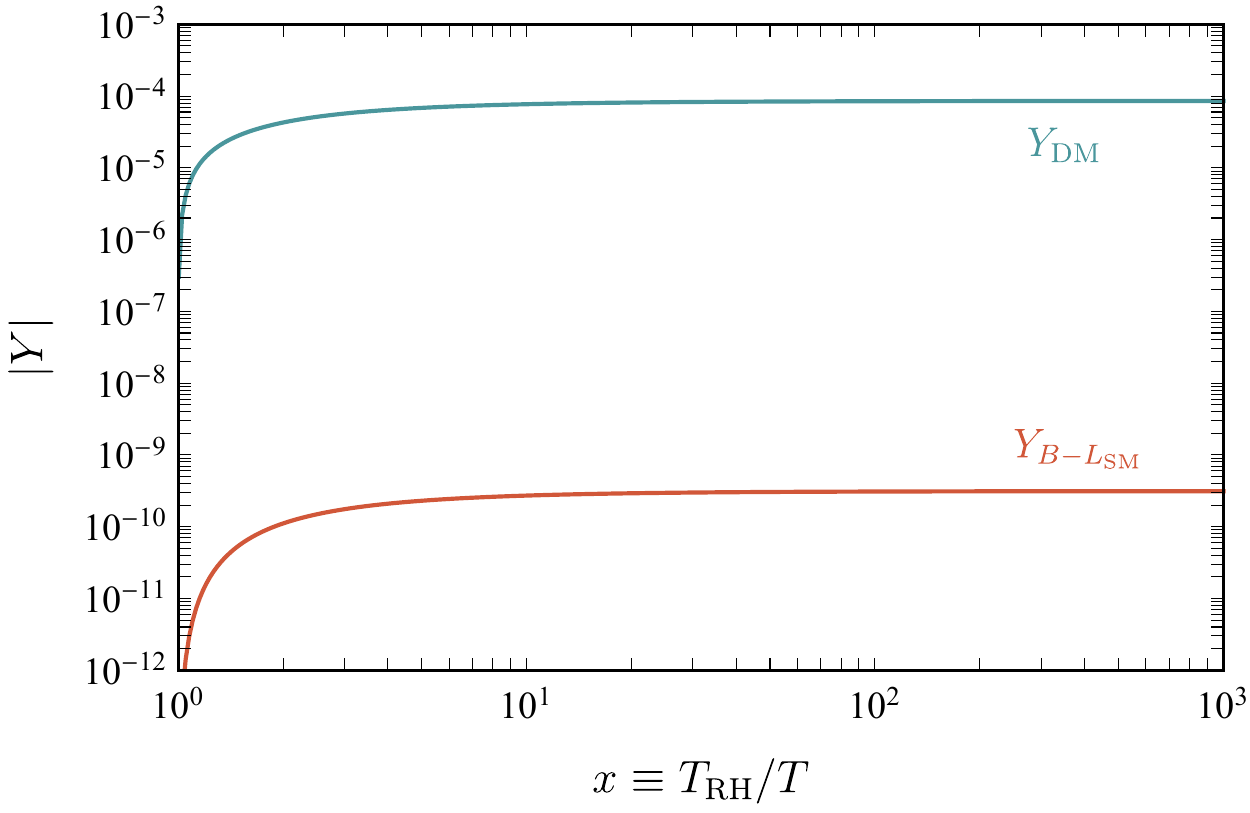}
      \caption{}
      \label{fig:DM_asymmetric_yields}
    \end{subfigure}
    \caption{(a) Tree-level and 1-loop Feynman diagrams of the dominant scattering processes that contribute to the generation of the DM abundance and the baryon asymmetry. (b) The yields $Y_{B-L_{\rm SM}}$ and $Y_{\rm DM}$ as functions of the dimensionless parameter $x\equiv T_{\rm RH}/T$ for a benchmark set of parameter values that reproduce the DM relic density abundance and matter-antimatter asymmetry of the universe.}
    \label{fig:UVFIBG_d5}
    \end{figure}

In this setup DM is produced at tree-level, mainly through scattering processes of the type $F_i\bar{e} \rightarrow \varphi\varphi$. The same process (along with, to a lesser extent, $F_i\bar{e} \leftrightarrow F_j\bar{e}$) also generates sector-wise lepton asymmetries in the SM and $F_i$ sectors through the interference between tree-level and one-loop diagrams depicted in Figure \ref{fig:Feyndiagrams}. The absolute values of the DM and asymmetric yields are presented in Figure \ref{fig:DM_asymmetric_yields} as a function of the dimensionless parameter $x\equiv T_{\rm RH}/T$. Once the electroweak sphalerons become active the SM sector asymmetry is partially converted it into a baryonic one, which freezes at sphaleron decoupling.

The predicted DM abundance is given by Eq.~\eqref{eq:YDMT} upon substituting $n=1$. The asymmetries generated in the $F$ and Standard Model sectors, on the other hand, are described by a set of three differential equations
\begin{subequations}
\begin{align}
-sHT\frac{\text{d}Y_{\Delta F_i}}{\text{d}T}&=-\big[F_i\bar{e}\leftrightarrow\varphi\varphi\big]-\big[F_i\bar{e}\leftrightarrow F_j\bar{e}\big]+(-1)^i\big[F_i\bar{F}_j\leftrightarrow e\bar{e}\big]\;,
\\\nonumber\\
-sHT\frac{\text{d}Y_{B-L_{\text{SM}}}}{\text{d}T}&=-\sum_i\big[F_i\bar{e}\leftrightarrow\varphi\varphi\big]\,,
\end{align}\label{eq:mastereqs}
\end{subequations}
where $Y_{\Delta F_i}\equiv Y_{F_i}-Y_{\bar{F}_i}$ and we have introduced the shorthand notations
\begin{subequations}\label{eq:notation conventions}
\begin{alignat}{2}
\left[a\,b\leftrightarrow c\,d\right]&\equiv\left(a\,b\leftrightarrow c\,d\right)-\left(\bar{a}\,\bar{b}\leftrightarrow \bar{c}\,\bar{d}\right)\,,
\\
\left(a\,b\leftrightarrow c\,d\right)&\equiv\frac{1}{S_{\text{in}}S_{\text{out}}}\int\text{d}\Pi_a \, \text{d}\Pi_b \, \text{d}\Pi_c \, \text{d}\Pi_d \, \left(2\pi\right)^4\delta^{(4)}\Big[\left|\mathcal{M}\right|^2_{ab\rightarrow cd}f_af_b\left(1\pm f_c\right)\left(1\pm f_d\right)\nonumber
\\
&\quad\qquad\qquad\qquad\qquad\qquad\qquad\qquad\qquad\qquad-\left|\mathcal{M}\right|^2_{cd\rightarrow ab}f_cf_d\left(1\pm f_a\right)\left(1\pm f_b\right)\Big]\,,
\end{alignat}
\end{subequations}
where $\delta^{\left(4\right)}$ is an abbreviation for $\delta^{\left(4\right)}\left(p_a+p_b-p_c-p_d\right)$ and $S_{\text{in}}$, $S_{\text{out}}$ are the symmetry factors for the incoming and outgoing states respectively, which are equal to $N!$ in case of identical particles and $1$ otherwise. The solution of these equations allows us to compute the predicted baryon asymmetry.

As shown in \cite{Goudelis:2022bls}, Eqs.~\eqref{eq:mastereqs} result in $Y_{\rm B-L_{\rm SM}}\sim \TRH^4/\Lambda^6$. From Eq.~\eqref{eq:YDMT}, on the other hand, setting $n=1$ one finds that the DM yield scales as $Y_{\rm DM}\sim \TRH/\Lambda^2$. This difference in the scaling  behaviour between $Y_{\rm B-L_{\rm SM}}$ and $Y_{\rm DM}$ is of critical importance for the analysis that follows.

All in all, the requirement for $\varphi$ to be a viable DM candidate can be satisfied simultaneously with that of achieving successful baryogenesis for $T_{\rm RH} > 10^{15}$ GeV and $\Lambda > 7\times 10^{15}$ GeV. The DM mass is then predicted to lie between $4 < m_\varphi < 15.5$ keV, with the lower bound coming from Lyman-$\alpha$ forest constraints.

\subsection{Dimension-6 operator}

The dimension-6 case works in full analogy with the dimension-5 one. In this case, the SM is extended by a pair of vector-like fermions $F_i$ which are charged under $U(1)_Y$, along with a Dirac fermion $\chi$ which acts as a DM candidate pair-created through a dimension-6 operator. The interaction Lagrangian that we consider is
\begin{equation}\label{eq:lagdim6}
\mathcal{L}\supset\frac{\lambda_{1}}{2\Lambda^2}\left(\bar{e}\PLH F_{1}\right)\left(\bar{\chi}\PRH\chi^c\right)+\frac{\lambda_{2}}{2\Lambda^2}\left(\bar{e}\PLH F_{2}\right)\left(\bar{\chi}\PRH\chi^c\right)+\frac{\kappa}{\Lambda^2}\left(\bar{e}\PLH F_{1}\right)\left(\bar{F}_2\PRH e\right)+\text{h.c.}\,,
\end{equation}
where for simplicity we ommited the $F_i$ gauge interaction terms.

The DM abundance can, again, be computed through Eq.~\eqref{eq:YDMT} by substituting $n=2$, whereas the evolution of the sector-wise lepton asymmetries is described by a system of equations which is identical with the one governing the dimension-5 case, with the substitution $\varphi \leftrightarrow \chi$.

In \cite{Goudelis:2022bls} it was found that the yields of ${B-L_{\rm SM}}$ and DM scale as $Y_{\rm B-L_{\rm SM}}\sim \TRH^8/\Lambda^{10}$ and $Y_{\rm DM}\sim \TRH^3/\Lambda^4$, respectively. The viable parameter ranges are found to be $T_{\rm RH} > 3\times 10^{13}$ GeV, $\Lambda > 3\times 10^{14}$ GeV and $4.6 < m_\chi < 18.1$ keV , \textit{i.e.} it is possible to satisfy all DM and baryogenesis constraints with relatively lower values of the reheating temperature compared to the dimension-5 case.

\section{Ultraviolet freeze-in baryogenesis in the presence of a decaying fluid}\label{sec:thefluid}

From the previous discussion it becomes clear that the mechanism of UV freeze-in baryogenesis is viable 
if the DM mass is small and the reheating temperature extremely large.
In the inflationary framework, thermalization occurs approximately when $\Gamma_\text{inf}\sim 3H$, where $\Gamma_\text{inf}$ is the decay rate of the inflaton, and the corresponding reheating temperature of the universe is 
\begin{equation}
T^\text{}_\text{RH} =\left(\frac{\pi^2}{10}g_{*\text{RH}}\right)^{-1/4} \sqrt{\Gamma_\text{inf} M_\text{Pl}}\,.
\end{equation}
Let us call $T_\text{max}$ the maximum possible reheating temperature of the universe  after the decay of the inflaton field. We  can write $T_\text{RH}= T_\text{max} \exp[{-\frac34 (1+\bar{w}_\text{RH}) \tilde{N}_\text{RH}}]$,
where $\bar{w}_\text{RH}$ and $\tilde{N}_\text{RH}$ are the mean equation of state and the e-folds during the reheating stage, respectively. The maximum temperature can be written in terms of the energy density at the end of inflation as $T_\text{max}=\rho^{1/4}_\text{end}(30/\pi^2g_{*\text{RH}})^{1/4}$ and is achieved in the instant reheating scenario, {\it i.e.} when $\tilde{N}_\text{RH}=0$.
From the hierarchy of the energy densities  $\rho_\text{end}  \lesssim V_* \simeq 3H_*^2 M^2_\text{Pl} 
    =({3\pi^2}/{2}) r_*{\cal P_R}(k_*) M^4_\text{Pl} $ 
we can obtain an upper bound on the reheating temperatures after inflation,
\begin{align}
 T_\text{RH} \lesssim   5 \times 10^{15} \text{GeV}  \left( \frac{106.75}{g_{*\rho}(T_\text{RH})}\right)^{1/4} \left( \frac{r_*}{0.036}\right)^{1/4} \left(\frac{{\cal P_R}(k_*)}{2 \times 10^{-9}} \right)^{1/4} \,,
\end{align}
where $g_{*\rho}(T_\text{RH})$ is the effective number of relativistic species upon thermalization, $r_*$  the tensor-to-scalar ratio and ${\cal P_R}(k_*)$ the curvature power spectrum  CMB scales, and we considered the BICEP2 bound $r_*<0.036$  \cite{BICEP:2021xfz}. 
This upper bound on the temperature restricts considerably the viable parameter space of our particle physics models of DM and baryon asymmetry generation. 

 However, it is possible  that after the inflaton decay the evolution of the universe could have been episodic with additional reheating events. 
In particular, if a late injection of entropy takes place during the pre-BBN cosmic evolution, the results obtained in the previous sections can be essentially modified as we demonstrate below.

\subsection{The fluid system of radiation and a scalar condensate}

In many theories beyond the SM there generically exist scalar fields with rather flat potentials and very weak or gravitationally suppressed interactions. These fields might never be in thermal contact with the plasma and decay slowly. 
Well-known examples are given by the  moduli fields, predicted by stringy setups, which may have important cosmological implications  \cite{Coughlan:1983ci, Ellis:1986zt, German:1986ds, deCarlos:1993wie, Banks:1995dt}. An early matter domination era has been studied in many different contexts, see \textit{e.g.} \cite{Fujii:2002fv, Erickcek:2011us, Easther:2013nga,  Dalianis:2018afb, Allahverdi:2021grt}. Let us call $X$ a scalar field of this sort. If the $X$-scalar potential is approximately quadratic about its minimum, it begins to oscillate when $m_X \sim H$, which corresponds to temperatures $T\sim (m_X M_{\rm Pl})^{1/2}$, and acts effectively as a pressureless scalar condensate. For $M_{\rm Pl}$-suppressed interactions the $X$ field decays at late times $\Gamma^{-1}_X\sim (M^2_{\rm Pl}/m^3_X)$ and for sufficiently large $m_X$ there is no conflict with the BBN predictions.
For typical initial displacements of the $X$ field an early matter domination era (EMD) can be realized. 

In particular, the energy density of a coherently oscillating scalar about the minimum of an effectively quadratic potential scales in the same way as pressureless matter, $\rho_X \propto a^{-3}$, and redshifts slower than the background radiation plasma. We suppose that at the cosmic time $t_{\text{dom},X} \gtrsim m_X^{-1}$ this scalar temporarily dominates the energy density of the early universe until it decays, at $t_{\text{dec},X}$, reheating the universe for a second time at the temperature  
 $T_{{\text{dec}},X}$. 
This transit EMD era dilutes any pre-existing abundances of the relativistic degrees of freedom by the amount 
\begin{equation} \label{DX}
\Delta _\text{EMD} \equiv \frac{S_\text{final}}{S_\text{initial}}  \approx \, \frac{T_{\text{dom},X}} {T_{\text{dec},X}}\,,
\end{equation}
where $S_\text{initial}$ and $S_\text{final}$ denote the comoving entropy of the universe at times well before ($t\ll t_{\text{dec},X}$) and after 
($t\gg t_{\text{dec},X}$)
 the decay of the X field.  
Note that $\Delta _\text{EMD}\approx (\rho_{\text{dom},X}/\rho_{\text{dec},X)})^{1/4}$. 

In order to incorporate this interfering process in our analysis, we assume a universe initially dominated by radiation where freeze-in DM and baryon asymmetry are produced. We also assume the presence of the aforementioned scalar field $X$. The $X$-decay enriches the radiation fluid and the evolution of the cosmological background is described by the  system of equations,
\begin{subequations}
    \begin{align} \
    \frac{d\rho_X}{d\tilde{N}} &\, =\,-3\rho_X -\frac{\Gamma_X}{H} \rho_X
 \\
 \frac{d\rho_\text{rad}}{d \tilde{N}} &\, =\,-4\rho_\text{rad}  + 
 (1-\text{B}_\text{DM})\frac{\Gamma_X}{H}\rho_X
 \,  
 \\
 \frac{d\rho_\text{DM}}{d \tilde{N}} &\, =\,-4\rho_\text{DM} +\text{B}_\text{DM}\frac{\Gamma_X}{H}\rho_X
 \,  
 \\
 \frac{dH}{d \tilde{N}}& \, = \,-\frac{1}{2H M^2_{\rm Pl}}\left(\rho_X+\frac{4}{3}\rho_\text{DM}+\frac{4}{3} \rho_\text{rad} \right)\,, \label{systH}
 \end{align}
\end{subequations}
 where  $d\tilde{N}=d(\ln a)=Hdt$ is the differential of the e-folds number.
 We  assume that $X$ decays symmetrically into SM particles and that the branching ratio of $X$ into DM particles, $\text{B}_\text{DM}\equiv \text{Br}(X\rightarrow \text{DM})$, is vanishing. For  $\text{B}_\text{DM}=0$ the above system can be effectively reduced to a system of two interacting fluids, the scalar field $X$ and the radiation. 
 Our DM particle is relativistic, albeit not in thermal equilibrium,  and its energy density only redshifts with time and is not sourced by the $X$ decay, $\text{B}_\text{DM} \Gamma_X \rho_X/H=0$. The result of the interchange in the energy density among the different components is that the early produced freeze-in DM abundance and lepton asymmetry are diluted by an amount $\Delta _\text{EMD}$ at later cosmic times $t>\Gamma_X^{-1}$.

 \begin{figure}
\centering
 \includegraphics [scale=.60, angle=0]{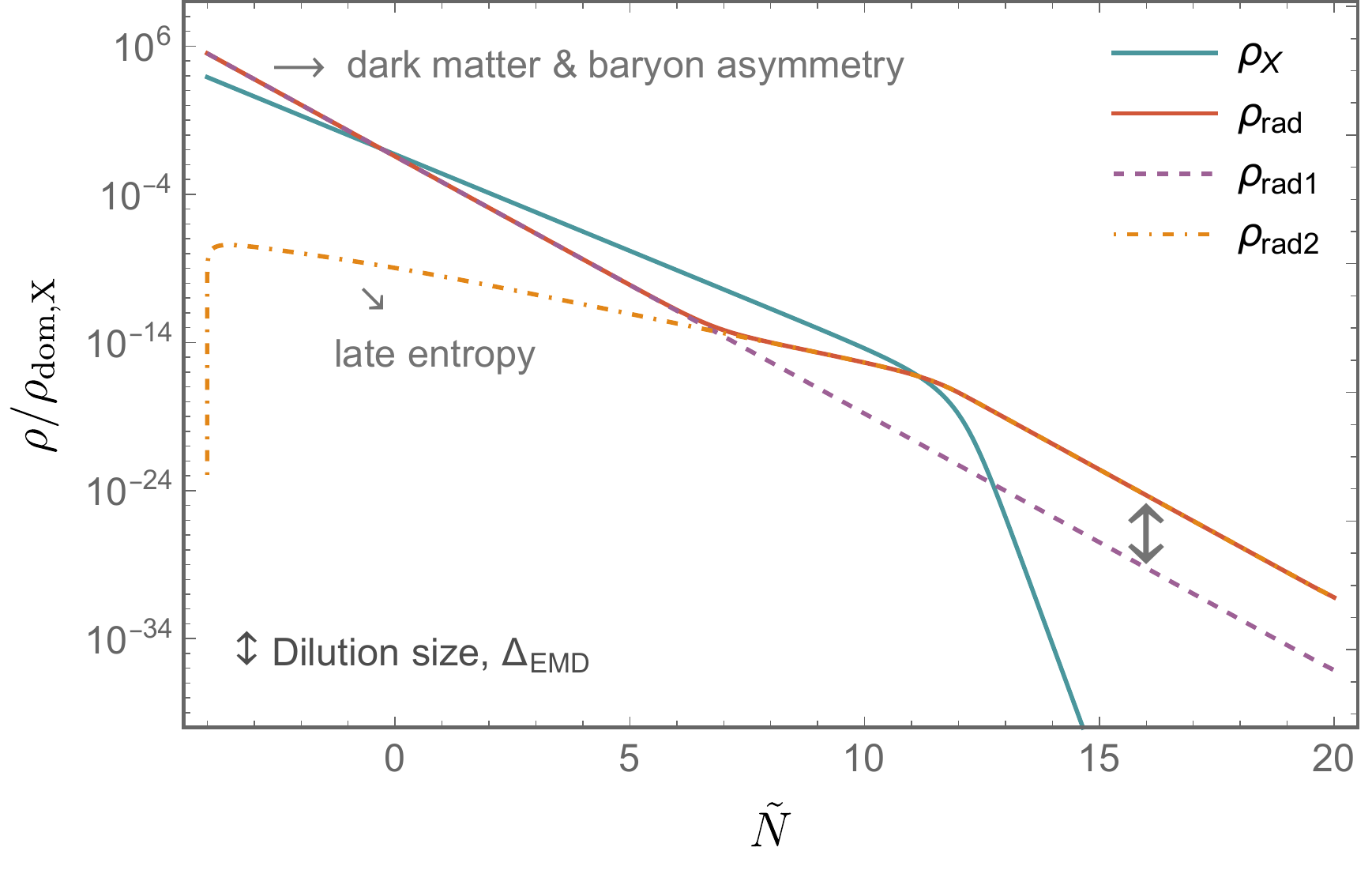} \,
\caption {\small {The evolution of the energy densities of 
two interacting fluids, radiation (red) and a scalar field $X$ (blue),
with respect to the e-folds  number $\tilde{N}$, normalized by the energy density at the moment of $X$ domination ($\tilde{N}=0$).
With a dashed line we depict the evolution of the radiation energy density  in the absence of $X$ decays  and with a dot-dashed the radiation component produced from the out of equilibrium $X$-decay.
The Figure illustrates a scenario where the scalar condensate domination lasts  $\tilde{N}_\text{EMD}\sim 12$ e-folds and realizes a dilution of size $\Delta _\text{EMD}\sim 10^4$.}}
\label{dens}
\end{figure}

 Figure~\ref{dens} depicts  the evolution of this two-fluid system for an illustrative scenario where the scalar $X$ dominates the energy density of the universe for about $\tilde{N}_\text{EMD} \sim 12$ e-folds  after choosing the value $\Gamma_X/H=10^{-8}$ for the scalar condensate fluid at the moment of early matter-radiation equality, that results in a dilution of size $\Delta _\text{EMD}\sim 10^4$. The $X$-condensate fluid is responsible for an early matter domination phase after the freeze-in processes of the DM and lepton asymmetry production ceased and when it decays away it reheats the universe for a second time.

\subsection{The viable parameter space in the scenario with late entropy production}

The viable parameter space of our microscopic models is, first of all, constrained by the inferred  present-day baryon asymmetry $Y_B = (8.71 \pm 0.06) \times 10^{-11}$ and the DM cosmic abundance $\Omega_{\rm DM} h^2 = 0.1200 \pm 0.0012$ \cite{Planck:2018vyg}. Furthermore, there are cosmological constraints on the DM mass value arising either from the possibility of DM thermalization or from large-scale structure formation considerations. Concretely, throughout our analysis we assume that the DM particles never reach thermal equilibrium with the SM thermal bath, whereas they are found to decouple from the plasma while they are still relativistic. 

Given these remarks, the non-thermalization requirement is satisfied if the number density of DM lies below its would-be equilibrium value during the period in which its interactions were sizeable. That is, 
for the period close to the reheating temperature we impose the constraint
\begin{equation}
Y_{\rm DM} < Y_{\rm DM}^{\rm eq}\,,
\label{eq:therm_bound_Y}
\end{equation}
otherwise the freeze-in assumption breaks down. Assuming that the DM particles saturate the entire DM content of the universe, \ie $\Omega h^2 \approx 2.8 \times 10^{8} \, Y_{\rm DM} \lrb{\mDM/\GeV}$, Eq.~\eqref{eq:therm_bound_Y} yields
\begin{equation}
\mDM > 0.17\,\injFact~\keV \,.
\label{eq:therm_bound}
\end{equation}

Structure formation bounds are related to the fact that the DM particles have a mass at the keV range and decouple when relativistic. Observations favour DM candidates that assist the formation of structures in the universe, which would be hampered if the free-streaming length of the DM particles is too large. The strongest constraints come from Lyman-$\alpha$ forest observations, which impose restrictions on the momentum of the particles created and on their redshift at relevant time-scales. Assuming a pure radiation-dominated universe, this imposes a lower bound on the freeze-in DM mass $\mDM \gtrsim m_{\rm Ly-\alpha}$, where $m_{\rm Ly-\alpha}\approx 4(4.6)$ keV in the $d=5(6)$ case \cite{Ballesteros:2020adh}. If, however, entropy is injected in the plasma, the momentum of the DM particles redshifts faster than the temperature of the plasma by a factor $\injFact^{1/3}$. We therefore expect the mass constraint to be relaxed by a similar factor\footnote{This estimate is corroborated by the findings of~\cite{Arias:2020qty}, in which a different method was employed.}. That is, the constraint becomes
\begin{equation}
\mDM \gtrsim m_{\rm Ly-\alpha}\,\injFact^{-1/3}\,.
\label{eq:mass_bound_UV}
\end{equation}
Note that by increasing the dilution size $\Delta_\text{EMD}$ the Lyman-$\alpha$ lower bound \eqref{eq:mass_bound_UV} on the DM mass becomes weaker, whereas the thermalization bound \eqref{eq:therm_bound} becomes stronger. For $\injFact \gtrsim 10\lrb{m_{\rm Ly-\alpha}/4\,\keV}^{3/4}$ the thermalization bound is the one that becomes the most restrictive one, both for the $d=5$ and for the $d=6$ case.

 Lastly, let us briefly comment on the fact that the decays of the surviving (post-freeze-out) $F_i$ particles should not disrupt the formation of light elements in the universe. Such BBN constraints, which essentially impose restrictions on the allowed combinations of $(m_F, \Lambda)$, were discussed in some detail in \cite{Goudelis:2022bls}. Given the fact that, as we will see in what follows, an early matter domination phase allows for lower values of $\Lambda$, \textit{i.e.} faster $F_i$ decay rates, this constraint is easily satisfied within the context discussed in the present paper.
\subsubsection{Dimension-5 operator and late entropy production}

In ~\cite{Goudelis:2022bls} it was shown that the viable parameter space found after a full numerical solution of Eqs.~\eqref{eq:mastereqs} could be determined to a very good approximation  by a few fairly simple analytical formulae. In the scenario that we consider here, in which entropy is injected in the plasma \textit{after} the generation of the DM and lepton asymmetry, these expressions are reformulated as
\begin{subequations}
\begin{align}
&\Lambda \approx 2 \times 10^{16}~\GeV \ 
\dfrac{\lrb{|\lambda_1|^2+|\lambda_2|^2}^2}{\sqrt{\left||\kappa||\lambda_1||\lambda_2|^2(|\lambda_1|^2-|\lambda_2|^2)\sin\lrb{\Delta \phi}  \right|}} 
\lrb{\dfrac{m_{\varphi}}{10~\keV}}^2 \, \injFact^{-3/2}
\\\nonumber\\
&T_{\rm RH} \approx 3 \times 10^{15} ~\GeV \ 
\dfrac{\lrb{|\lambda_1|^2+|\lambda_2|^2}^3}{\left||\kappa||\lambda_1||\lambda_2|^2(|\lambda_1|^2-|\lambda_2|^2)\sin\lrb{\Delta \phi}  \right|} 
\lrb{\dfrac{m_{\varphi}}{10~\keV}}^3 \, \injFact^{-2}  \, ,
\end{align} \label{eq:d5_obs}
\end{subequations}
where $\Delta\phi\equiv \phi_1-\phi_2-\phi_3$ and $\phi_1,\phi_2,\phi_3$ are the phases of the couplings $\lambda_1,\lambda_2$ and $\kappa$, respectively. Since the parameter $\injFact$ can take values within a range that spans many orders of magnitude, we expect a significant modification of the viable parameter space compared to the case of pure radiation domination. Alternatively, we can treat $\TRH$ as free and constrain $\mDM$, using $m_{\varphi}=\Omega_{\rm DM}\rho_c/(Y_{\varphi}(T_0)s_0)$ and Eq.~\eqref{eq:YDMT} with $n=1$, as 
\begin{equation}
    m_{\varphi} \approx  10~\keV \left(\dfrac{\Lambda}{2 \times 10^{16}~\GeV }\right)^2 \left(\dfrac{3 \times 10^{15}~\GeV}{\TRH}\right) \dfrac{\injFact}{\lrb{|\lambda_1|^2+|\lambda_2|^2}} \,.
\label{eq:mDM_d5_obs}
\end{equation}

    \begin{figure}[t]
    \centering
    \begin{subfigure}{.5\linewidth}
      \includegraphics[width=\linewidth]{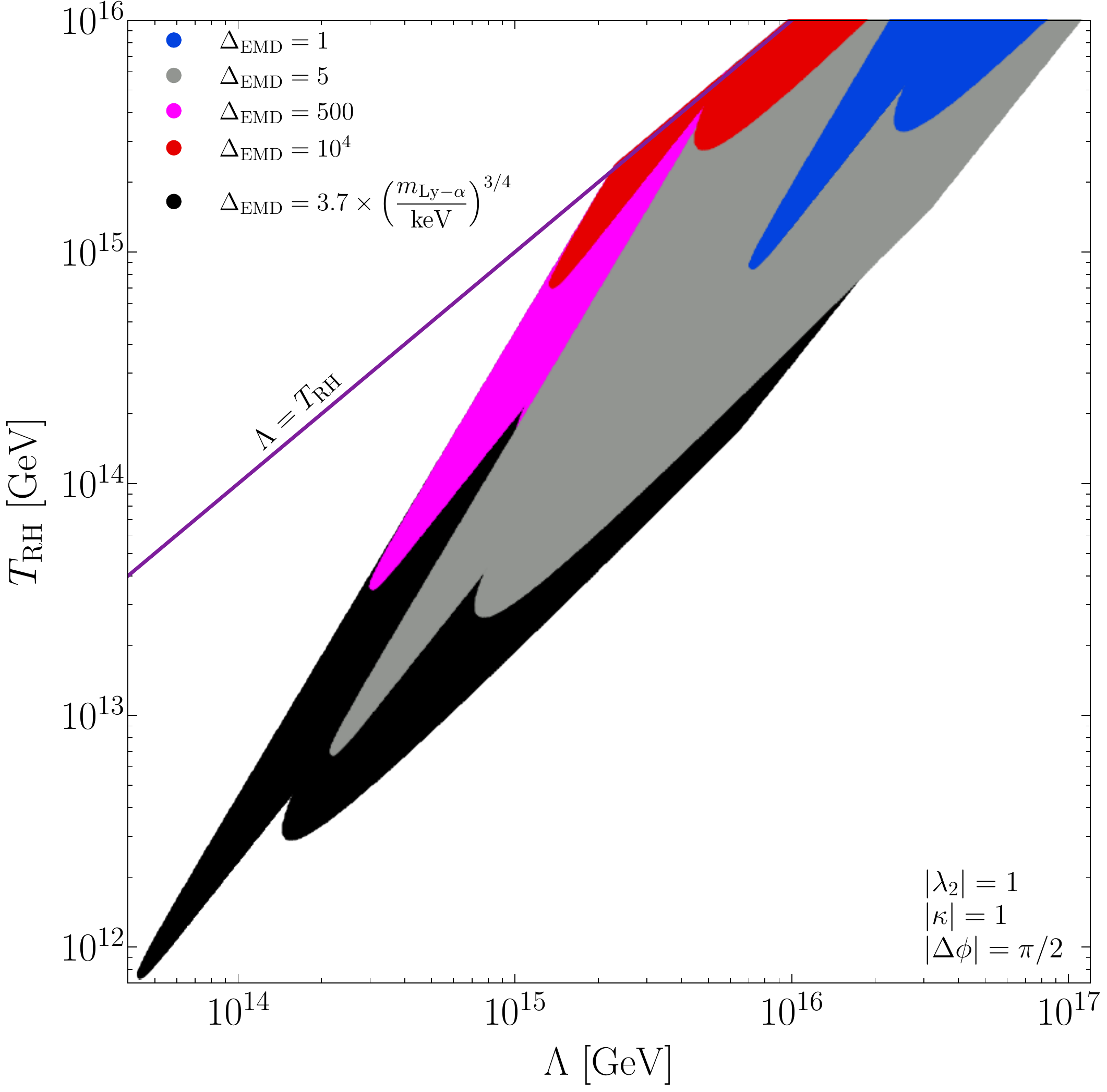}
      \caption{}
      \label{fig:TRH_d5_Lam}
    \end{subfigure}\hspace*{\fill}
    \begin{subfigure}{.5\linewidth} 
    \includegraphics[width=\linewidth]{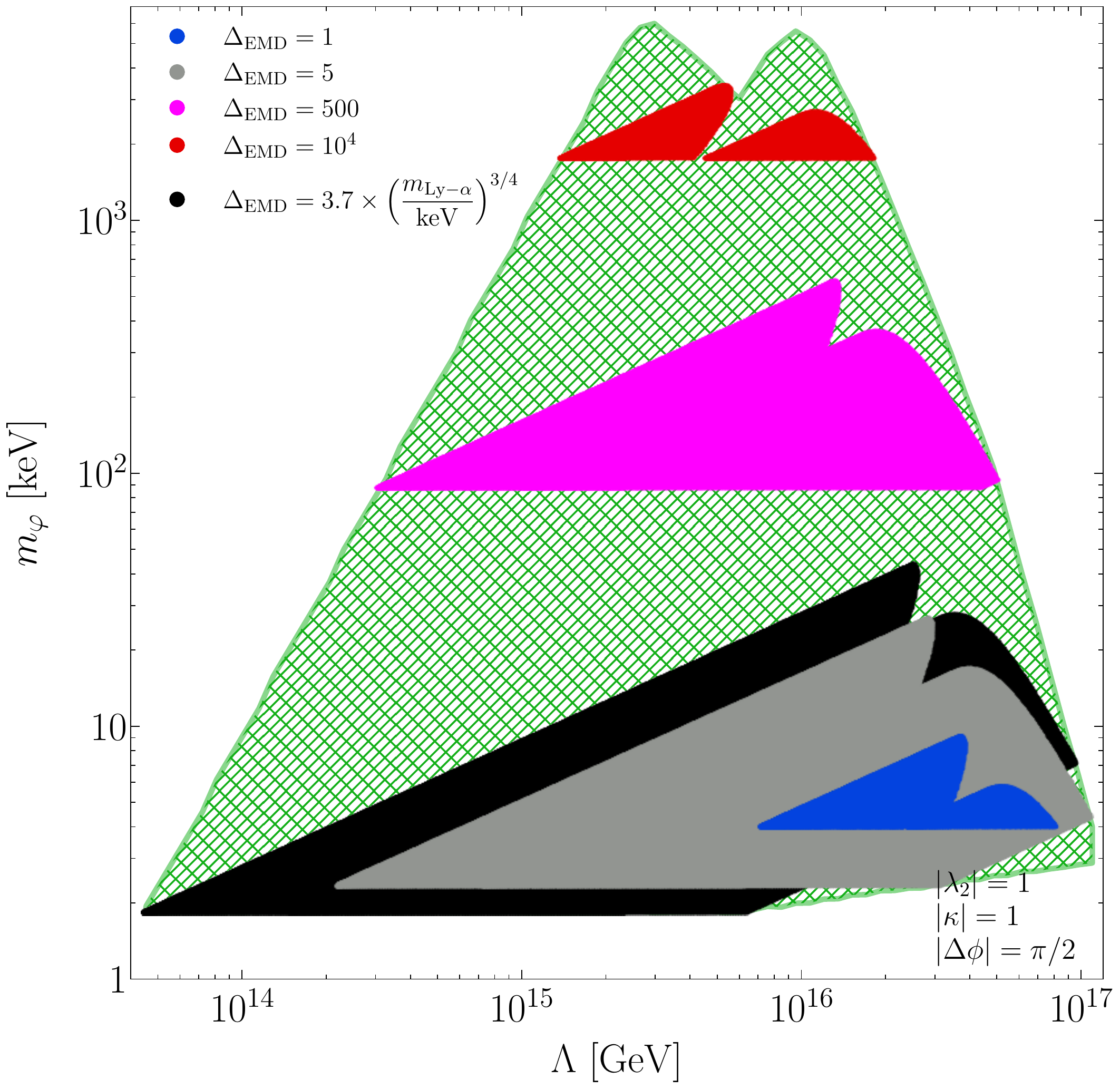}
      \caption{}
      \label{fig:mDM_d5_Lam}
    \end{subfigure}
    \caption{
        {(a)} The allowed parameter space for the $d=5$ model in the $(\Lambda, \TRH)$ plane. The parameter space is saturated for $\injFact \approx 3.7\left(m_{\rm Ly-\alpha}/\rm keV\right)^{3/4}$ due to the thermalization of DM; \ie as more entropy is injected in the plasma, the parameter space starts to be pushed towards the limiting line $\TRH = \Lambda$ and becomes smaller. This results in an absolute upper bound $\injFact \lesssim 4\times 10^4$ (for $|\lambda_2| = 1$). 
        {(b)} The allowed parameter space in the $(\Lambda,m_{\varphi})$ plane. Here, larger values of $\injFact$ lead to a decrease of the allowed area, but for increased values of the DM mass. The green hatched area indicates the full parameter space allowed for all values of $\injFact$.
        }
    \label{fig:scan_d5}
    \end{figure}

Our results are presented in Figure~\ref{fig:scan_d5}. In the left-hand side panel, we show the viable parameter space in the $(\Lambda, T_{\rm RH})$ plane for the benchmark choice of parameters $\left| \lambda_2 \right| = 1$, $\left| \kappa \right| = 1$ and $\left| \Delta \phi \right| = \pi/2$, with all other model parameters being varied freely. The different colour regions correspond to different values for the dilution factor $\injFact$, whereas the solid line depicts the limit $T_{\rm RH} < \Lambda$, above which the use of Effective Field Theory becomes unjustified and knowledge of the full UV completion of our theory would be required. In the right-hand side panel, our results are projected on the $(\Lambda, m_{\varphi})$ plane.

 We see that the available parameter space turns out to behave fairly non-trivially for increasing values of the dilution parameter $\injFact$. As expected, assuming no entropy is injected in the plasma ($\injFact = 1$, blue region), our results match the ones presented in \cite{Goudelis:2022bls}. As $\injFact$ increases to values $5$, $500$ and $10^4$ (gray, magenta and red regions, respectively), lower values of $\Lambda$ (or alternatively, higher values of the DM mass) and $T_{\rm RH}$ become possible, since DM can be largely overproduced initially and subsequently diluted. Interestingly, however, this does not happen in a monotonous manner (\textit{i.e.} each subsequent region is not just broader than the former one). This is due to the interplay of two factors: first, and focusing on the left-hand side panel of Figure~\ref{fig:scan_d5}, it is important to keep in mind that the asymmetry generated in the SM lepton sector \textit{also} undergoes dilution and, hence, for a fixed value of $T_{\rm RH}$, lower values of $\Lambda$ are required for successful baryogenesis. As this trend continues for increasing $\injFact$, $\Lambda$ starts approaching the limit of $T_{\rm RH}$, which is a limiting value for our EFT-based analysis. In other words, the allowed parameter space in the $(\Lambda, T_{\rm RH})$ plane gets squeezed towards line $T_{\rm RH} = \Lambda$. 
 
 The allowed values of the DM mass, on the other hand, behave much more as expected: keeping all other parameters constant, larger dilution factors can be compensated by larger values of the DM mass in order to saturate the {\it Planck} data bounds. A large part  of the available parameter space corresponds to rather small values for the  dilution factor, which is an outcome of the DM non-thermalization condition. Note that the green hatched area in Figure~\ref{fig:mDM_d5_Lam} indicates the full available parameter space once $\injFact$ is allowed to vary continuously between $1$ and $\sim 10^4$.
 
 In summary, for moderate ${\cal{O}}(1-10^2)$ values of $\injFact$ the allowed parameter space increases whereas for much higher values $\gtrsim{\cal{O}}( 10^4)$ one starts approaching the limitations of an EFT-based analysis. The DM mass is always allowed, or even required, to take larger values with respect to the case of a standard thermal history.
\subsubsection{Dimension-6 operator and late entropy production}

In the case of a dimension-6 operator, when a late entropy production takes place, the corresponding expressions that constrain  the energy scale of the EFT and the reheating temperature are reformulated as follows, 
\begin{subequations}
\begin{align}
& \Lambda \approx 7.8 \times 10^{14}~\GeV \ 
\dfrac{\lrb{|\lambda_1|^2+|\lambda_2|^2}^4}{\lrb{\left||\kappa||\lambda_1||\lambda_2|^2(|\lambda_1|^2-|\lambda_2|^2)\sin\lrb{\Delta \phi}  \right|}^{3/2}} 
\lrb{\dfrac{m_{\chi}}{10~\keV}}^4 \ \injFact^{-5/2}
\\\nonumber\\
&T_{\rm RH} \approx 7 \times 10^{13} ~\GeV \ 
\dfrac{\lrb{|\lambda_1|^2+|\lambda_2|^2}^5}{\left||\kappa||\lambda_1||\lambda_2|(|\lambda_1|^2-|\lambda_2|^2)\sin\lrb{\Delta \phi}  \right|^2} 
\lrb{\dfrac{m_{\chi}}{10~\keV}}^5 \ \injFact^{-3} \,.
\end{align} \label{eq:d6_obs}
\end{subequations}
Similarly to the $d=5$ case, the value of the DM mass, $m_{\chi}$, can be expressed in terms of $\Delta_{\rm EMD}$ as 
\begin{equation}
    m_{\chi} \approx 10~\keV \ \lrb{\dfrac{\Lambda}{7.8 \times 10^{14}~\GeV}}^4 \lrb{\dfrac{7 \times 10^{13} ~\GeV }{\TRH}}^3 \dfrac{\injFact}{\lrb{|\lambda_1|^2+|\lambda_2|^2}} \,.
\label{eq:mDM_d6_obs}
\end{equation}

\begin{figure}[t]
    \centering
    \begin{subfigure}{.5\linewidth}
      \includegraphics[width=\linewidth]{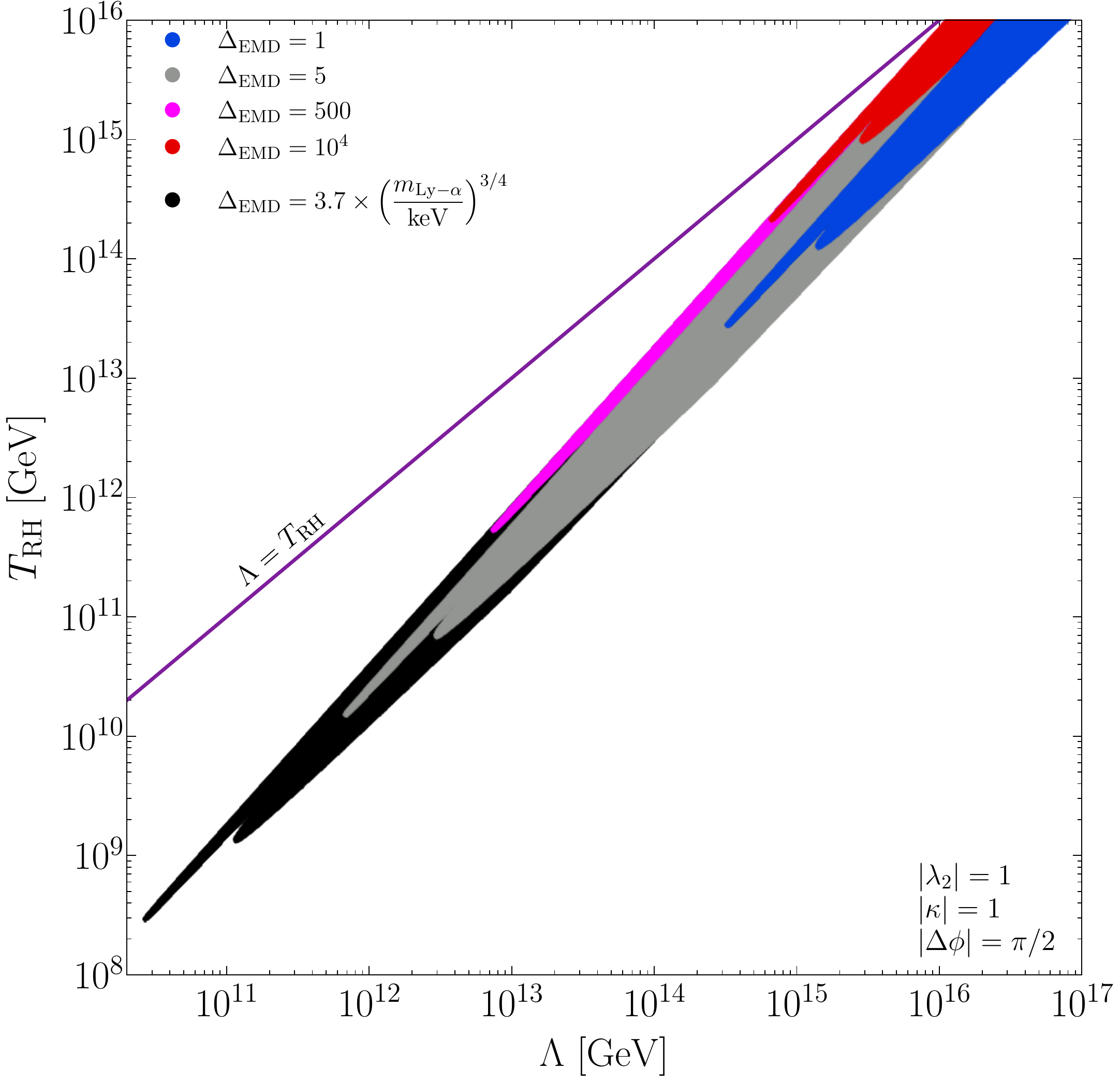}
      \caption{}
      \label{fig:TRH_d6_Lam}
    \end{subfigure}\hspace*{\fill}
    \begin{subfigure}{.5\linewidth}
    \includegraphics[width=\linewidth]{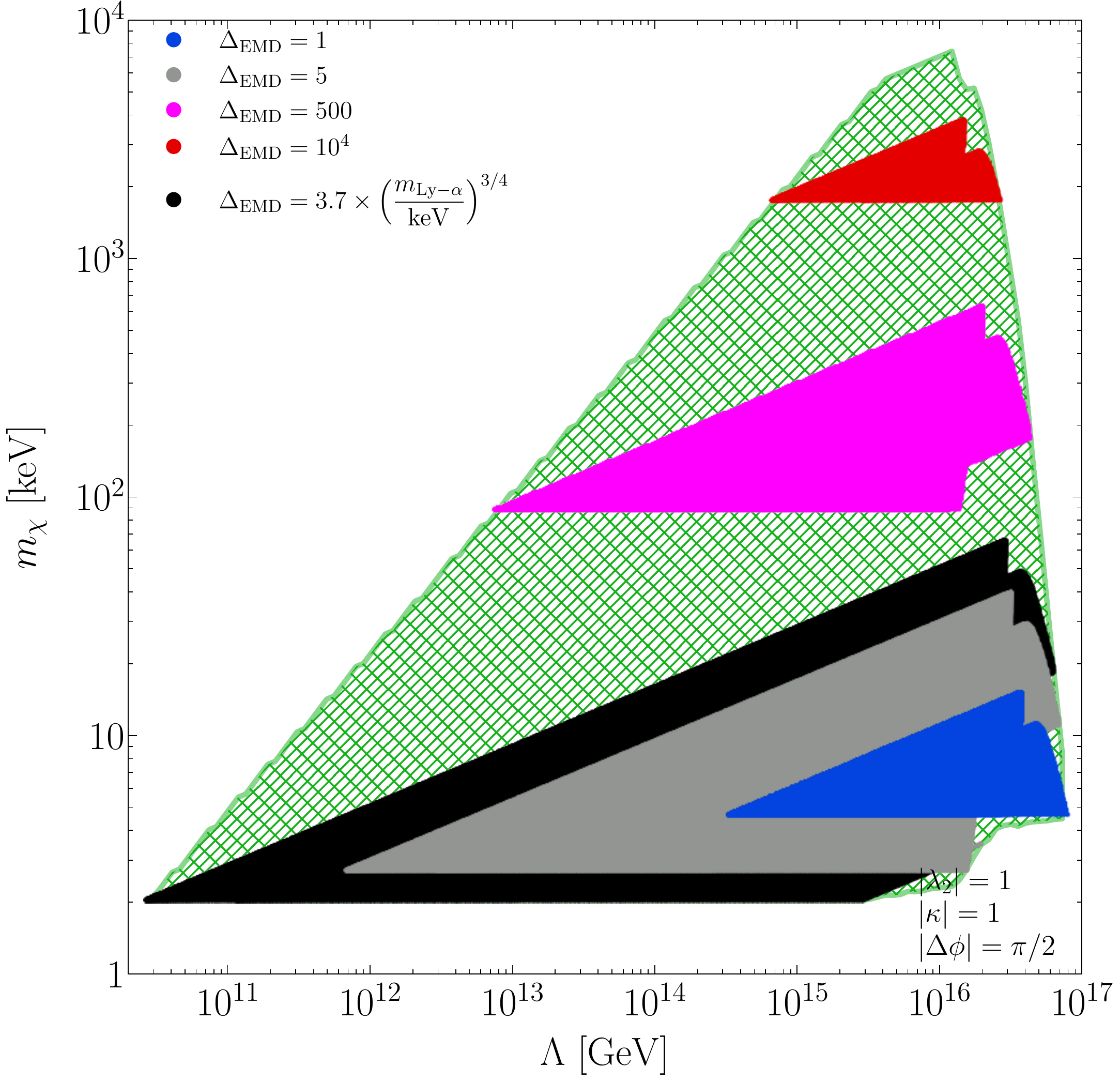}
      \caption{}
      \label{fig:mDM_d6_Lam}
    \end{subfigure}
    \caption{The viable parameter space for the $d=6$ case, projected on the (a) $(\Lambda,T_{\rm RH})$ and (b) $(\Lambda,{\chi})$ planes. All colour codes are identical to the ones in Figure \ref{fig:scan_d5}.}
    \label{fig:scan_d6}
    \end{figure}

Our results are presented in Figure~\ref{fig:scan_d6}. Once again, in the left-hand side panel we show the viable parameter space in the $(\Lambda, T_{\rm RH})$ plane for the benchmark choice of parameters $\left| \lambda_2 \right| = 1$, $\left| \kappa \right| = 1$ and $\left| \Delta \phi \right| = \pi/2$. Again, the different colour regions correspond to different values for the dilution factor $\injFact$, whereas the solid line depicts the limit $T_{\rm RH} < \Lambda$. In the right-hand side panel our results are projected on the $(\Lambda, m_{\chi})$ plane.

The general qualitative trend is identical to the dimension-5 case, with the allowed $(\Lambda, T_{\rm RH})$ parameter space increasing for moderate values of the dilution factor and, eventually, getting squeezed towards the limit $\Lambda = T_{\rm RH}$. Higher values of the DM mass become available, or even necessary, depending on the exact value of $\injFact$.


\section{The parameter space of our EFTs and the inflationary observables}
\label{sec:observables}

As we saw, our two models for the simultaneous generation of the baryon asymmetry and dark matter are sensitive to the cosmic thermal history. They require rather large initial temperatures for the thermal plasma, $T_\text{RH} \gtrsim 10^{13}$ GeV, in the case of pure radiation domination, albeit smaller temperatures, $T_\text{RH} \gtrsim 10^{8}$ GeV, become possible in the case of dilution through late entropy injection. The precise range of the allowed temperatures changes with the dimension of the effective operator. This observation is extremely important because, as we will show in what follows, once supplemented with a specific inflationary hypothesis the two cases can yield distinct predictions in terms of inflationary observables.

\subsection{Inflationary e-folds and the the dilution size}

Inflation is the leading paradigm for the explanation of the origin of the primordial density perturbations that grew into the CMB anisotropies. The {\it Planck} mission has constrained the spectral index of the two-point correlation function of the primordial scalar perturbation to be $n_s=0.9649\pm 0.0042$ (68\% CL) and the tensor-to-scalar ratio to $r<0.1$ \cite{Planck:2018jri}. More recently BICEP2 presented the bound $r<0.036$ (95\% CL) at $k=0.05$ Mpc$^{-1}$ \cite{BICEP:2021xfz} and, future  
 CMB-Stage 4 experiments \cite{ CMB-S4:2020lpa} will have the sensitivity to detect a $r>0.003$ signal. Additionally, experiments such as EUCLID \cite{EuclidTheoryWorkingGroup:2012gxx} and cosmic 21-cm surveys \cite{Mao:2008ug, Pritchard:2011xb} will have the capacity to achieve a precision of $10^{-3}$ in the value of $n_s$ and further constrict the existing observational bounds. 
 
 In this era of precision cosmology, the dependence of the inflationary predictions on the number of e-folds becomes increasingly important. The fact that the e-folds number is associated with the details of the reheating stage implies that the value of reheating temperature of the universe can, potentially, be  tested \cite{Kinney:2005in, Martin:2010kz, Easther:2013nga, Dai:2014jja, Munoz:2014eqa,  Drewes:2015coa, Gong:2015qha, Dalianis:2018afb}. 
The reheating temperature is a crucial parameter for UV-sensitive freeze-in DM and baryogenesis scenarios such as the ones we examine here. 
For these reasons,  a study of DM production and baryogenesis in conjunction with the inflationary dynamics can be a powerful strategy to explore  
the physics of the early universe \cite{Allahverdi:2020bys}.

By specifying the reheating stage the uncertainty in the number of  e-folds is raised and the inflationary predictions are localized around a spot on the ($n_s, r$) contour.
Therefore, benchmark $T_\text{RH}$ values required for our models to be viable can be tested together with specific inflationary models. Let us recall the basic expressions that relate the inflationary observables with  the reheating stage. The observable number of e-folds $N_*$ is defined as 
\begin{align}
 N_*\equiv \int^{t_\text{end}}_{t_*} \text{d}t H =\ln(a_\text{end}/a_*)\,,
 \end{align}
where $t_*$ denotes the moment that the observable (pivot) scale exits the Hubble radius and $t_{\rm end}$ the moment that inflation ends. Inflation theory predicts a relation between $N_*$  and the postinflationary number of e-folds. Assuming a single reheating stage, one finds the  expression \cite{Liddle:2003as, Planck:2018jri}
\begin{align} \label{N*th}
    N_*
    \approx 67 -\ln\left( \frac{k_*}{a_0 H_0}\right) +\frac14 \ln\left( \frac{V^2_*}{M^4_\text{Pl} \rho_{\text{end}}}\right) -\frac{1-3\bar{w}_\text{RH}}{4} \tilde{N}_\text{RH} -\frac{1}{12} \ln \left(g_{*\text{RH}} \right)\,, 
\end{align}
where $\tilde{N}_\text{RH}=\ln(a_\text{RH}/a_\text{end})=\ln(\rho_\text{end}/\rho_\text{RH})/(3+3\bar{w}_\text{RH})$
 stands for the e-folds number of the post-inflationary reheating period and $\bar{w}_\text{RH} = \left\langle p\right\rangle/\left\langle \rho\right\rangle$ is the averaged effective equation of state for a coherently oscillating inflaton  field,  given by the ratio of the average pressure over the average energy density \cite{Shtanov:1994ce}.

The transient scalar domination era introduced in Section \ref{sec:thefluid} alters the standard cosmological scenario of a smooth and continuous radiation domination era that follows inflation. In particular, it
modifies the relation between the observable CMB pivot scale $k_*$ 
and the comoving curvature scale of our present universe as
\begin{equation} \label{kmod}
    \frac{k_*}{a_{0} H_{0}} = \frac{a_*}{a_\text{end}} \frac{a_\text{end}}{a_\text{RH}}
    a_\text{RH}
\left(    \frac{1}{a_{\text{dom},X}}
\frac{a_{\text{dom},X}}{a_{\text{dec},X}}
\frac{a_{\text{dec},X}}{1} \right)
\frac{1}{a_\text{eq}}
\frac{a_\text{eq}H_\text{eq}}{a_{0}H_0}\frac{H_*}{H_\text{eq}}\,,
\end{equation}
where the subscripts 
refer to the epoch of the radiation-matter equality (eq) and the present time (0). The parenthesis in the relation (\ref{kmod})
is the modification introduced by the transient scalar domination phase and changes the $N_*$ relation  (\ref{N*th}) as follows,
\begin{equation} \label{Nx}
    \frac{1-3\bar{w}_\text{RH}}{4} \tilde{N}_\text{RH}\, \longrightarrow \,\,
     \,\, \frac{1-3\bar{w}_\text{RH}}{4} \tilde{N}_\text{RH} +  \frac{1}{4} \tilde{N}_\text{EMD}\,,
\end{equation}
where  $\tilde{N}_\text{EMD}\equiv \ln(a_{\text{dec},X}/a_{\text{dom},X})$. We assume  $\bar{w}_X\approx 0$ for the scalar field domination phase, \textit{i.e.} a scalar condensate, and so $\tilde{N}_\text{EMD} \approx \frac{1}{3} \ln\left( {\rho_{\text{dom},X}}/{\rho_{\text{dec},X}}\right)\approx \frac43\ln({T_{\text{dom},X}}/ {T_{\text{dec},X}})$ modulo a possible change in the thermalized degrees of freedom. 
 Therefore, the dilution size (\ref{DX}) has an exponential dependence on the duration of the transit scalar domination,
\begin{equation} \label{Dx}
    \Delta _\text{EMD}\approx \exp{\left(\frac{3}{4} \tilde{N}_\text{EMD} \right)}\,.
\end{equation}
What is of particular interest is that
 the transient scalar domination induces a shift of the  $N_*$ value by the amount $\delta N_* \approx \tilde{N}_\text{EMD}/4$,
\begin{equation} \label{N*}
N_*=N^{(\text{th})}-
\delta N_* \, \approx \, N^{(\text{th})}-\frac{1}{3}\ln(\Delta _\text{EMD})\,.
\end{equation}
This shift is understood as a deflection from a reference value that we call {\it thermal reference} value  $N^{(\text{th})}$, that is the e-folds number value  if there were no late entropy production after the inflaton decay.
Taking, now, into account the effect of the dilution the inflationary e-folds are given by the expression, 
\begin{equation} \label{Nall}
  N_*  \approx 67 -\ln\left( \frac{k_*}{a_0 H_0}\right) +\frac14 \ln\left( \frac{V^2_*}{M^4_\text{Pl} \rho_{\text{end}}^{1+\gamma_\text{RH}}}\right)
  +\gamma_\text{RH} \ln\left(\frac{T_\text{RH}}{\Delta_\text{EMD}^{1/(3\gamma_\text{RH})}} \right),
\end{equation}
where $\gamma_\text{RH}\equiv (1-3\bar{w}_\text{RH})/(3+3\bar{w}_\text{RH})$ and we have omitted the minute 
$\gamma_\text{RH} \ln(\pi^2/30)/4$ and $(3\gamma_\text{RH}-1)\ln(g_{*\text{RH}})/12$ terms. A common assumption for the equation of state during reheating is $\bar{w}_\text{RH}=0$, that corresponds to $\gamma_\text{RH}=1/3$.

For the two models presented in Sec.~\ref{sec:thefluid}, the reheating temperature can be expressed as a function of the parameters of each model and the dilution size. We get respectively,
\begin{itemize}
    \item Dimension-5 operator:
     \begin{align}
      N_*= N_{123}+
      \gamma_\text{RH} \ln\left(\frac{F_5(\lambda_1, \lambda_2, \kappa, \Delta \phi)}{\Delta_\text{EMD}^{2+1/(3\gamma_\text{RH})}} \right) 
      +3\gamma_\text{RH} \ln\left(\frac{m_{\varphi}}{10 \text{keV}} \right)\,,
     \end{align}
    \item Dimension-6 operator:
     \begin{align}
     N_*= N_{123}+
      \gamma_\text{RH} \ln\left(\frac{F_6(\lambda_1, \lambda_2, \kappa, \Delta \phi)}{\Delta_\text{EMD}^{3+1/(3\gamma_\text{RH})}} \right) 
      +5\gamma_\text{RH} \ln\left(\frac{m_{\chi}}{10 \text{keV}} \right) \,,
     \end{align}
\end{itemize}
where $N_{123}$ stands for the sum of the first three terms in the rhs of Eq.~(\ref{Nall}) and 
 its numerical value can be found after substituting $k_*=0.002 \,  \text{Mpc}^{-1}$, $H_0\approx 1.4\times 10^{-33}$ eV and $\ln(10^{10}A_s)=3.089$. 
The functions $F_5$ and $F_6$ can be read off Eqs.~(\ref{eq:d5_obs}) and (\ref{eq:d6_obs}), respectively, and give the $T_\text{RH}$ value in GeV for $m_{\varphi(\chi)}=10$ keV and $\Delta_\text{EMD}=1$.
The above equations express the inflationary e-folds numbers as a function of the underlying inflation dynamics and the dilution size $\Delta_\text{EMD}$. Also, there is  
 an implicit dependence on the parameters of each microscopic model.
 In the simple case where the parameters $\lambda_1$, $\lambda_2$ and $\kappa$ had fixed values, $N_*$ would depend only on the dilution size and for a specific inflationary model an analytic result could be obtained.

\begin{figure}[t]
    \centering
    \begin{subfigure}{.5\linewidth}
      \includegraphics[width=\linewidth]{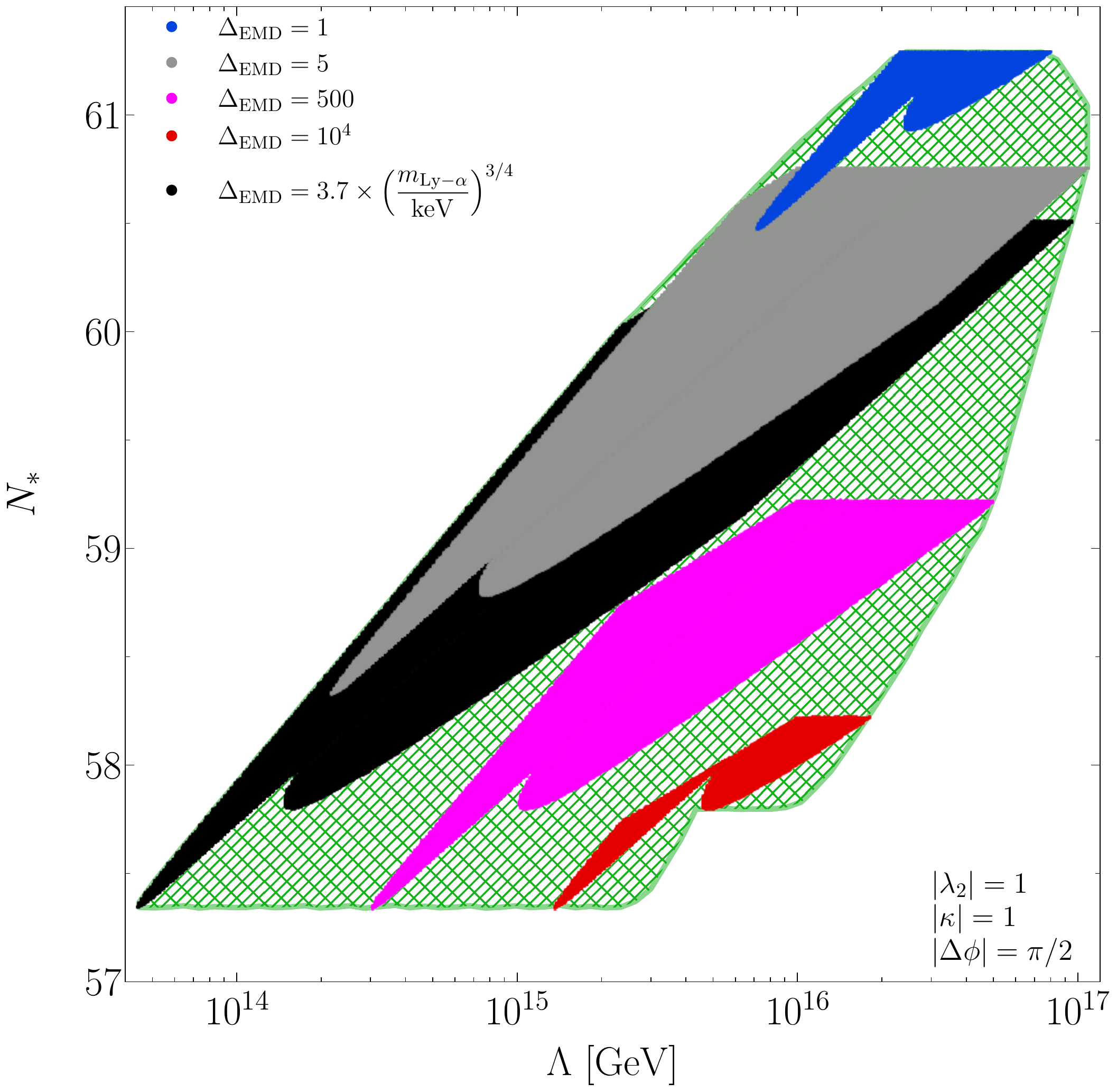}
      \caption{}
      \label{fig:N*_d5_Lam}
    \end{subfigure}\hspace*{\fill}
    \begin{subfigure}{.5\linewidth}
    \includegraphics[width=\linewidth]{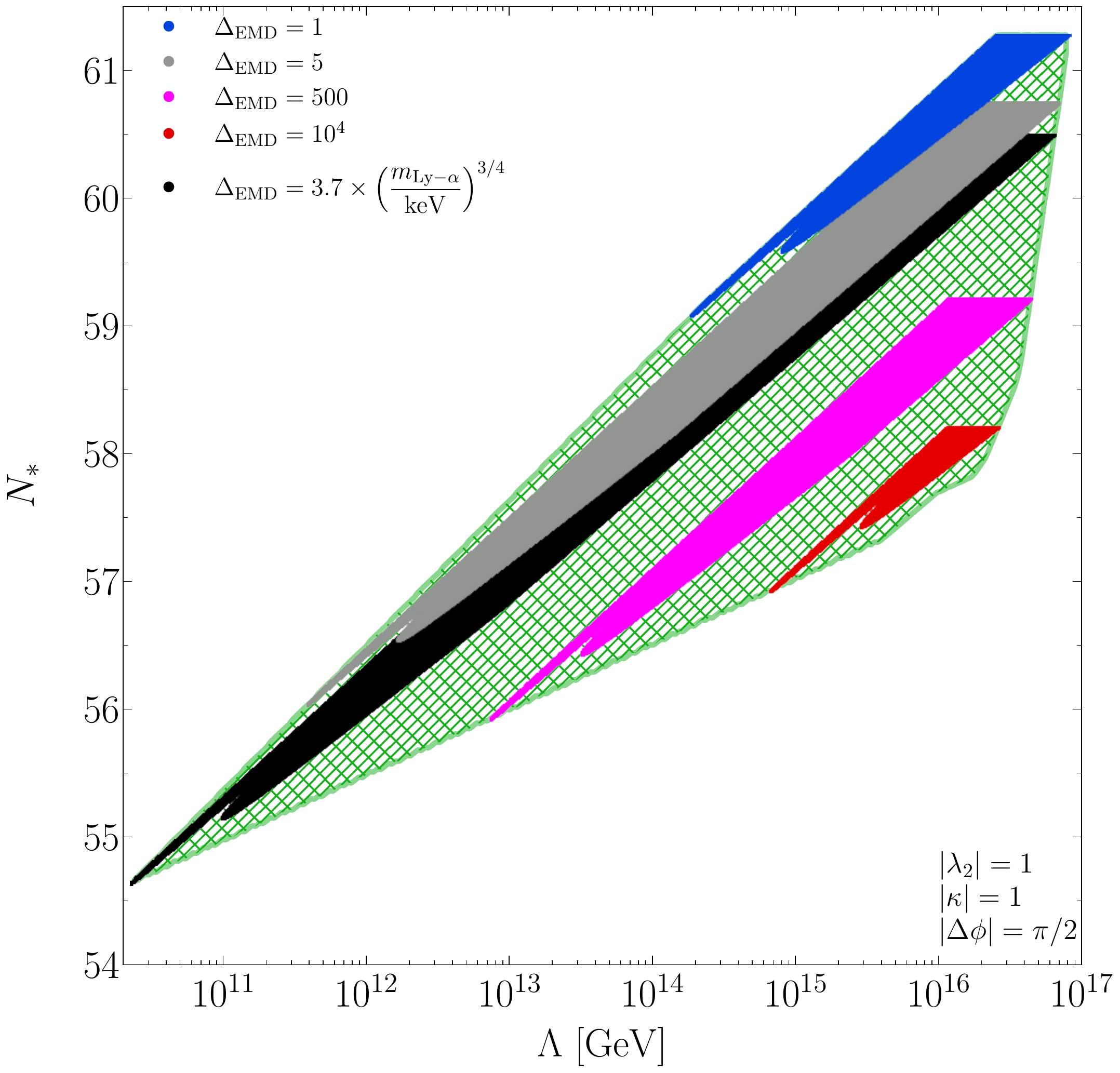}
      \caption{}
      \label{fig:mDM_N*_d5_Lam}
    \end{subfigure}
    \caption{ 
         The $\Lambda-N_*$ relation for the case of a dimension-5 (a) and dimension-6 (b) EFT operator assuming fiducial values for the underlying inflationary parameters. The green hatched region indicates, again, the full available parameter space. 
         }
    \label{fig:LambdaNstar}
    \end{figure}

In Figure~\ref{fig:LambdaNstar} we show the available parameter space in the $(\Lambda,N_*)$ plane, for the dimension 5 and 6 operators respectively, after performing a scan of the viable parameter space for  benchmark $\lambda_2$, $\kappa$ and $\Delta \phi$ values and allowing $\lambda_1$ to vary. We have also assumed  fiducial values for the relevant parameters of the underlying inflationary dynamics, $r_*=10^{-3}$,  $V_*/\rho_\text{end}=10$ and $\bar{w}_\text{RH}=0$. A ballpark relation for the $\Lambda-N_*$ scaling  is $N_* \sim \log\Lambda +44$ with $N_* > 57$ for the case of a dimension-5 operator and $N_* > 54$ for dimension-6 case. Remarkably, we see that an inflationary observable can in principle constrain the dimensionality of the underlying EFT operator responsible for generation of DM and the baryon asymmetry of the universe, at least for the minimal cases considered here.

A more specific  $\Lambda-N_*$ parameter space can be obtained if a particular inflationary model is chosen. By doing so we can test in more detail which inflationary model
matches best with our microscopic EFT constructions. 
In the following, and trying to be as  concise as possible,  we proceed with a generic  illustrative analysis to examine the implications of our EFT models to the predictions of some commonly discussed inflationary potentials. 
With this combined analysis a connection   with CMB observables is made possible.

\subsection{The shift in $n_s$ and $r$ due to late entropy production}

The  $N_*$ by itself is an observable quantity only through a particular $n_s=n_s(N)$ relation. 
At leading order the scale dependence of the spectral index  can be written as  
\begin{equation} \label{expansion}
n_s(k) \approx 1 -\frac{{\alpha}(N)}{N}+ \frac{\beta(N)}{N^2}\,.
\end{equation}
However, there is no common expression for the functions ${\alpha}(N) $ and  $\beta(N)$ and the possible shift of the spectral index can be tested only within specific models of inflation. 
The exact expression for the spectral index value is given by the Hubble-flow parameters, $\epsilon_{n+1}={\text{d} \ln |\epsilon_n|}/{\text{d}N}$,
and at leading order it is 
\begin{equation}
    n_s=1-2\epsilon_{1*} -\epsilon_{2*}+ {\cal O}(\epsilon^2)\,.
\end{equation}
 The spectral index shift is $\delta n_s \approx \alpha_s \delta N $, where ${\alpha_s}$ is the running of the spectral index and in terms of the dilution size it reads,
\begin{equation} \label{nsa}
    \delta n_s\approx \frac{\alpha_s}{3}\ln(\Delta_\text{EMD})\,.
\end{equation}
This expression depends on the Hubble-flow parameters and differs from one model to another.
Generically, though, the minimal inflationary models typically predict
$\alpha_s>0$ (or $\alpha(N)>0$ in Eq.~(\ref{expansion})), which means that the spectrum tilt becomes more red when dilution of the radiation plasma takes place. Let us also point out that there is also a shift in the tensor-to-scalar ratio, albeit the tensor spectrum has not been detected yet.

Next, we consider specific models of inflation in order to quantify and demonstrate the impact of our simple microphysical scenarios on the inflationary observables. We choose four minimal and effectively single field  models of infation which are  representative enough for our purposes and at the same time capture the main characteristics of several concrete models.

\subsubsection{Specific models of inflation} \label{InfModels}

Each inflation model yields a particular $n_s(N)$ expression. Here we will express in an informative way the shift in the spectral index, $\delta n_s$, due to a pre-BBN dilution of the thermal plasma, for four broad classes of models. We will use Eq.~(\ref{nsa}) that gives the difference from the spectral index thermal value, $n_s^{(\text{th})}=n_s(N^{(\text{th})})$, where
$N^{(\text{th})}$ depends on the reheating temperature and on the intrinsic inflationary dynamics. The $\delta n_s$ value is a measure of the EMD duration and enables us to select the parameter space of our EFT  and vice versa. 
 For a favourable reheating temperature and inflationary model the $n_s^\text{(th)}$ value can be calculated and the difference from the observable value  $\delta n_s$ implies a $\Delta_\text{EMD}$ which in turn, 
 according to the information that can be read off from Figures~\ref{fig:TRH_d5_Lam} and \ref{fig:TRH_d6_Lam}, selects  
  a value $\Lambda$ for our EFT operators. Of course this argument can also be used the other way round and, assuming specific underlying particle dynamics, concrete inflationary models can be favored or disfavored.
In the rest of this subsection we will set $M_{\rm Pl}=1$.\\

\noindent 
{\it Large field models}.
They are characterized by a single monomial potential $ V(\phi)=M^{4-p} \phi^p$
 where $p$ is a positive number \cite{Linde:1983gd, Kawasaki:2000yn, McAllister:2008hb}. The {\it Planck} data disfavor this sort of inflationary models unless $p\lesssim 1$. 
The first and second Hubble-flow parameters read, $\epsilon_{1*}=p/(4N_*+p)$, $\epsilon_{2*}=1/(N_*+p/4)$ and from Eq.~(\ref{nsa}) the $n_s(N)$ relation is obtained with $\alpha_s \approx (2+p)/(2N^2)$. According to Eq.~(\ref{DX}) it is $\tilde{N}_\text{EMD}\approx\frac43 \ln(\Delta _\text{EMD})$ and $\delta N_*\approx -\text{N}_\text{EMD}/4$ hence, for this class of models, we find the spectral index shift  
\begin{align} \label{LF}
    \delta n_s \approx 
    -\frac43\, 10^{-4}\left(1+\frac{p}{2}\right)\left(\frac{50}{N^{(\text{th})}(T_\text{RH})}\right)^2 \ln(\Delta _\text{EMD})
\end{align}
from the thermal value $n_s^\text{(th)}=(4N^\text{(th)}-p-4)/(4N^\text{(th)}+p)$. 
From Eqs.~(\ref{N*th}) and (\ref{nsa}), and for concrete values of $p$, $\bar{w}_\text{RH}$ and $T_\text{RH}$, the values of $N^\text{(th)}$ and $n_s^\text{(th)}$ can be computed.\\

\noindent
{\it Small field models}.
This is the class of potentials in which inflation starts from small field values $\varphi \ll M_\text{Pl}$.
The most characteristic example, called also {\it hilltop}, is when the inflaton is rolling away from an unstable maximum of the potential \cite{Linde:1981mu, Albrecht:1982wi}. The typical potential is
    $V(\phi)=M^{4}\left[1-\left(\frac{\phi}{\mu} \right)^p\right]$.
The first and second Hubble-flow parameters  are $\epsilon_{*1}=\frac{p^2}{2\mu^2}(N_* p(p-2)/\mu^2)^{-2(p-1)/(p-2)}$ and $\epsilon_{*2}={2}(p-1)/((p-2)N_*)$. We consider values for the parameters $\mu\lesssim M_\text{Pl}$ and $p>2$. 
It holds that $\epsilon_{*1}\ll\epsilon_{*2}$, that is, the tensor modes are much more suppressed compared to large field models. The running of the spectral index is 
$\alpha_s\approx 2N_*^{-2}(p-1)/(p-2)$, hence, for these models the spectral index shift from the thermal value is 
\begin{align} \label{SF}
    \delta n_s \approx 
    -\frac83  10^{-4}\left(1+\frac{1}{p-2}\right)\left(\frac{50}{N^{(\text{th})}(T_\text{RH})}\right)^2 \ln(\Delta _\text{EMD}) \,,
\end{align}
where $n_s^\text{(th)}\approx1-2(p-1)/((p-2)N^\text{(th)})$.

Besides, negative $p$ values also amount to phenomenologically and theoretically well-motivated inflationary potentials; see \cite{Dvali:2001fw, Martin:2013tda} and references therein. Contrary to hilltop models here the field rolls-down from large to small $\phi$ values and such setups are frequently referred to as {\it inverse hilltop} models.
The first and second Hubble-flow parameters, the running and the shift of the spectral index can be read off the previous formulas, with the replacement $p\rightarrow -p$. Accordingly, the thermal value for the spectral index is $n_s^\text{(th)}\approx
1-2(p+1)/((p+2)N^\text{(th)})$ which is larger than the $n_s^\text{(th)}$ predicted by hilltop models for the same $N^\text{(th)}$ number.\\

\noindent
{\it $R^2$-type/$\alpha$-attractors plateau potentials}.
A potential fully consistent with the {\it Planck} data constraints has a plateau with a mild slope given by the general expression $V(\phi)=M^{4}\big(1- e^{-\sqrt{{2}/({3\alpha_E}})\,\, \phi }\big)^{2n_E}$.
This potential is the $\alpha$-attractors $E$-model in the Einstein frame \cite{Kallosh:2013hoa, Kallosh:2015lwa}.
For $n_E=1$ and $\alpha_E=-1/9$ in the inflaton potential we get the Linde-Goncharov model \cite{Goncharov:1983mw}.
The case $n_E=1$ and $\alpha_E=1$ is known as the Starobinsky or $R^2$ potential \cite{Starobinsky:1980te}, where $\phi$ is a gravitational scalar. It can be also derived in the framework of different  theories, as for example $\alpha$-attractors \cite{Kallosh:2013hoa, Kallosh:2013yoa} and Higgs inflation  \cite{Bezrukov:2007ep} in which $\phi$ has unsuppressed interactions and decay rate.
At leading order the first  Hubble-flow parameter is $\epsilon_{1*}=(3/4)/N^2_*$ and  
the running of the spectral index is 
$\alpha_s=2/N^2-(0.11+3\ln(N))/N^3$.
 Hence   for the Starobinsky-type plateau of models the spectral index shift is at leading order
\begin{align} \label{attr}
    \delta n_s \approx
    -\frac{8}{3}\, 10^{-4}\left(\frac{50}{N^{(\text{th})}(T_\text{RH})}\right)^2 \ln(\Delta _\text{EMD}) 
\end{align}
from the thermal value $n_s^\text{(th)}\approx
1-2/N^\text{(th)}+(0.81+1.5\ln(N^\text{(th)}))/(N^\text{(th)})^2$. The logarithmic correction to $\delta n_s$ is one order of magnitude smaller and, thus, neglected here. 

The above expressions help us to demonstrate at a quantitative level that the combined examination of inflation together with a given microscopic model for DM and baryon asymmetry generation (in our case, through minimal dynamics described by $d=5$ and $d=6$ operators), potentially supplemented by an early matter domination epoch, provides a promising strategy for model selection.

\subsubsection{Observational implications}

The observational uncertainty of the early state of the the universe for  temperatures $T\gtrsim 5$ MeV $\sim T_\text{BBN}$ \cite{Kawasaki:1999na} implies a significant uncertainty on the inflationary e-folds of size $\delta N_* \lesssim 15$. This uncertainty is partitioned between  $\tilde{N}_\text{RH}$ and $\tilde{N}_\text{EMD}$, see Eq.~(\ref{Nx}). In our models, the former is quite small because of the large reheating temperatures required. It is $T_\text{RH} \gtrsim 10^{12}\,\text{GeV}$ for a dimension-5 operator and $T_\text{RH}\gtrsim 10^{8}\,\text{GeV}$ for a dimension-6 operator, thus the uncertainty in $\tilde{N}_\text{RH}$ is constrained to be  $(1- 3\bar{w}_\text{RH})\tilde{N}_\text{RH}/4 \lesssim 2-4$. The latter uncertainty is associated with the dilution size. Typically, our models require a dilution size of order $\Delta _\text{EMD} \lesssim 10^4$ and we expect a distinct shift in the e-folds number $\delta N_* \lesssim 4$.
 
In principle, for a complete and predictable inflationary model the mean equation of state during reheating and the inflaton decay rate can be estimated and, thus, the uncertainty in $ \tilde{N}_\text{RH}$ is lifted. The most notable example is that of the Starobinsky inflation model, where the inflaton is a homogeneous condensate of scalar gravitons after the end of the inflationary expansion. It decays with Planck-suppressed interactions and the e-folds number is estimated without uncertainty to be $N_*=N^{\text{(th)}}=54$. Hence, the thermal index value is  $n_s^{(\text{th})}=0.965$. The predictability of the Starobinsky model makes it ideal for combined studies with DM/baryogenesis production scenarios. This has been done in Ref. \cite{Dalianis:2018afb} in the framework of the supergravity Starobinsky model.

Note that the inflationary models that we presented in subsection (\ref{InfModels})
are incomplete in terms of reheating: the large field monomial modes   (\ref{LF}) for $p\lesssim 1$ do not feature a smooth potential minimum, while the hilltop and inverse hilltop models also require a low-energy completion of their potential in order to end inflation. 
We assume here that during reheating the inflaton experiences a quadratic potential 
which is the lowest order term in a Taylor expansion about the origin, hence $\bar{w}_\text{RH}\approx 0$. 
Reheating lasts for a period roughly $\Gamma_\text{inf}^{-1}$ where $\Gamma_\text{inf}$ is the inflaton decay rate.

\begin{figure}[t]
    \centering
    \begin{subfigure}{.49 \linewidth}
      \includegraphics[width=\linewidth]{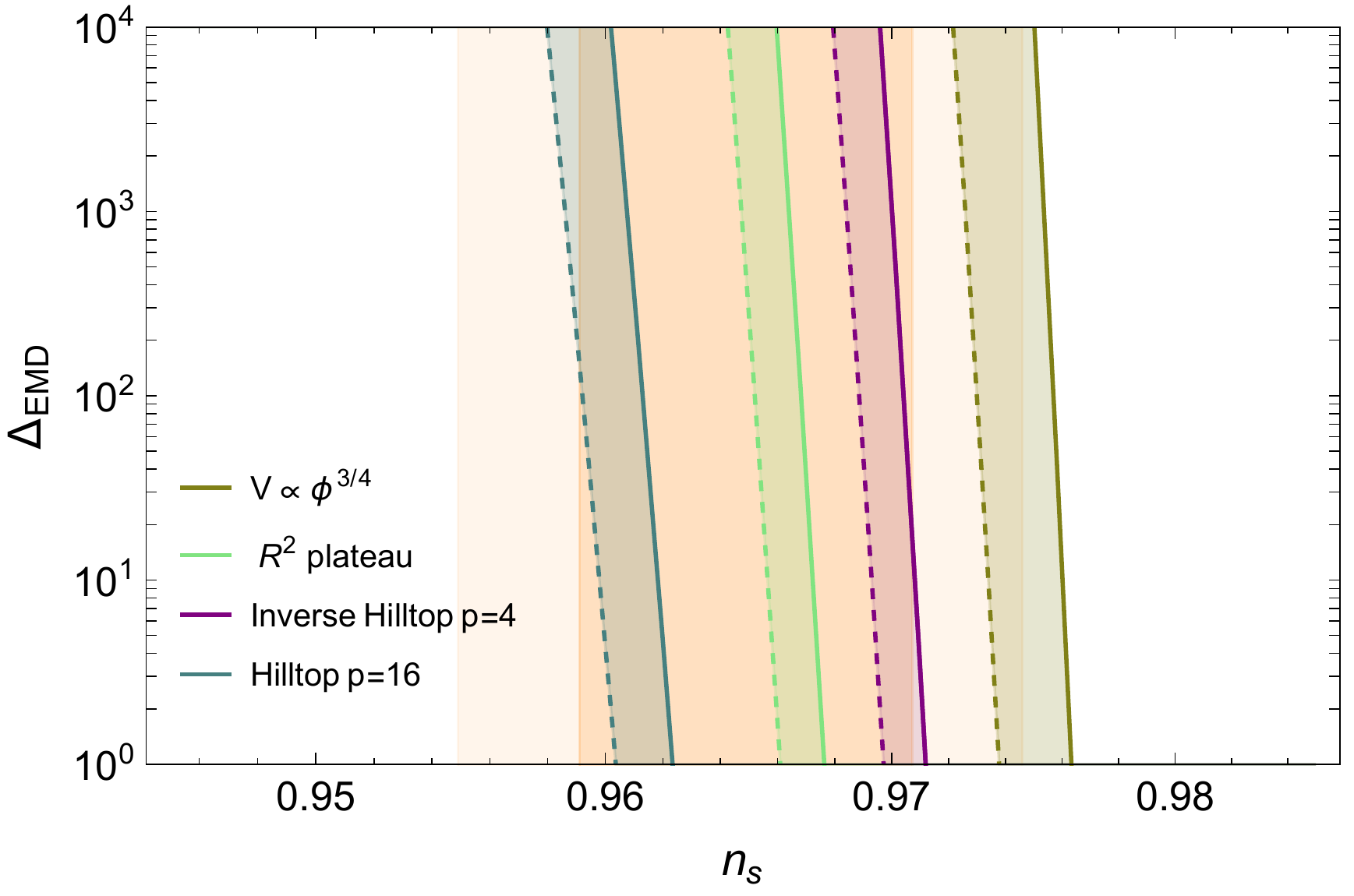}
      \caption{}
      \label{fig:ns4}
    \end{subfigure}\hspace*{\fill}
    \begin{subfigure}{.49\linewidth}
    \includegraphics[width=\linewidth]{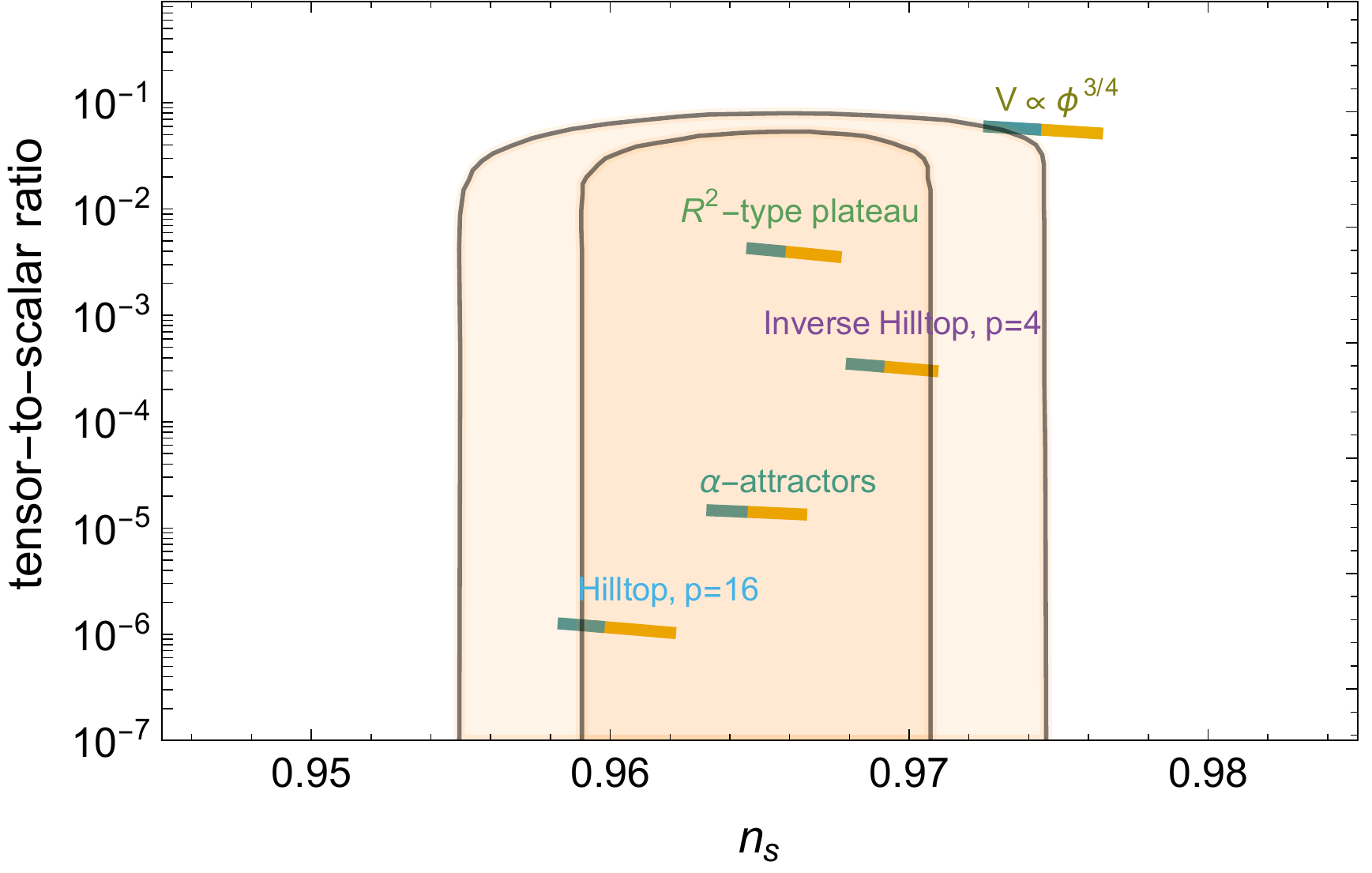}
      \caption{}
      \label{fig:ns5}
    \end{subfigure}
    \caption{ 
     { (a)} The $n_s$ value for four types of inflationary models and  a dilution $\Delta _\text{EMD}\leq 10^{4}$. The
      reheating temperature range is that of the $d=5$ case, $10^{12}<T_\text{RH}<5\times 10^{15}$ (solid and dashed lines respectively).
 { (b)}  The range of the allowed $(n_s, r)$ values for five types of inflationary models when combined with our $d=5$ and $d=6$ EFT models. The blue part segment is compatible only with the $d=6$ case corresponding to a fermionic DM candidate, while the orange segment with both the $d=5$ and $d=6$ scenarios.
  The pink colored contours are the CL regions from {\it Planck} results \cite{Planck:2018jri}.
}
    \label{fig:nsPlot}
    \end{figure}

In Figure~\ref{fig:ns4} the $n_s$ value for four types of inflationary models and for  the $d=5$ case is depicted when a dilution of size $\Delta_\text{EMD} \leq 10^4$ takes place. 
For $\Delta_\text{EMD}=1$ we recover the $n_s^\text{(th)}$ value, while for $\Delta_\text{EMD}>1$ there is a $\delta n_s$ shift according to Eqs.~(\ref{LF}), (\ref{SF}) and (\ref{attr}). The size of each band is determined by the range of the reheating temperatures, 
 $10^{12}\, \text{GeV} \lesssim \, T_\text{RH} \, \lesssim \, 5\times 10^{15}$ GeV. 
 Note that not all the area of the band is permitted and Figure~\ref{fig:N*_d5_Lam} must be read alongside Figure~\ref{fig:ns4}. 
 For a benchmark reheating temperature value $T_\text{RH}$ 
 the $n_s^\text{(th)}$ can be calculated, while for a benchmark $\Lambda$ value a $\Delta_\text{EMD}$ value is selected and, together, from the information contained in Figure~\ref{fig:ns4} the $n_s$ value can be inferred. Hence, when the results depicted in Figure~\ref{fig:N*_d5_Lam} are combined with those of Figure~\ref{fig:ns4}, the energy scale  of our EFT can be interrelated with the $n_s$ observable. A similar analysis is performed for the $d=6$ case as well.

A concrete observational implication of our analysis is that, according to the findings presented in Figure~\ref{fig:LambdaNstar},
the $d=5$ and $d=6$ cases 
can give rise to distinct predictions, at least for some part of the available parameter space,
since the $d=5$ case is not compatible with a fiducial inflationary model that yields $N_*$ below a certain value whereas for the $d=6$ case lower values are possible. 
Therefore, the CMB data can indeed inform microscopic realizations of UV-dominated freeze-in baryogenesis once these are supplemented by a concrete inflationary hypothesis.

 In Figure~\ref{fig:ns5} 
 we depict the corresponding range of the possible $n_s$ and $r$ values (colored segments) implied by our microscopic models against the contours  from {\it Planck} data in combination with BICEP2/Keck Array BK14 data \cite{Planck:2018jri}. The orange part of the segments is compatible with both the $d=5$ and $d=6$ EFT models, whereas the blue part is compatible only with the $d=6$ case. In particular, for a large field model with $p=3/4$ the $d=5$ case requires approximately $N_*\gtrsim 55$ whereas the one $d=6$ requires $N_*\gtrsim 52$. Accordingly, for the Starobinsky-type  plateau we get $N_*\gtrsim 56$ ($d=5$) and $N_*\gtrsim  54$ ($d=6$), for the inverse hilltop model with $\mu=0.1 M_{\rm Pl}$ and $p=4$  $N_*\gtrsim 55$ ($d=5$) and $N_*\gtrsim 53$ ($d=6$), for the hilltop model with  $\mu=0.1 M_{\rm Pl}$ and $p=16$ $N_*\gtrsim 54$ ($d=5$) and $N_*\gtrsim 52$ ($d=6$), and for an $\alpha$-attractor model with $r=10^{-5}$ we obtain $N_*\gtrsim 56$ ($d=5$) and $N_* \gtrsim 54$ ($d=6$).
 
 Our findings show that models described by large field inflation are in better agreement with CMB constraints when combined with our $d=6$ EFT scenario, whereas models such as the hilltop match better with the $d=5$ case.  With future observations reaching a higher level of accuracy, it might be possible to discriminate further between different inflationary models and probe the details of the pre-BBN era. 
 
\section{Summary and conclusions}\label{sec:conclusions}

In this paper we studied the implications of an early matter-dominated cosmological era for a mechanism that allows for the simultaneous generation of dark matter and a viable matter-antimatter asymmetry in the universe through the same microscopic out-of-equilibrium (freeze-in) processes. We considered two examples, in which these processes are described by simple non-renormalizable operators: in the first case the DM is a scalar produced through a dimension-5 operator, while in the second case it is a fermion produced through a dimension-6 one. In both scenarios the freeze-in processes occur in the ultraviolet, \textit{i.e.} at very high temperatures, well above the BBN energy scale. Assuming a standard cosmic thermal history, according to which inflation was followed by an uninterrupted radiation-dominated era, both scenarios require a very high reheating temperature and light dark matter, close to the Lyman-$\alpha$ structure formation bounds. 

However, it is uncertain whether the universe was continuously dominated by radiation since the end of inflation and until the BBN epoch. For example, in the presence of exotic massive non-relativistic particles that decay slowly, it is possible that a period of non-standard cosmology was realized. A typical example is that of an early matter domination era (EMD) that can alter the cosmological abundances of particle species and asymmetries, once entropy is injected in the plasma upon the decay of the exotic species. In our freeze-in scenario the relic DM abundance and the size of baryon asymmetry have a different scaling with temperature, therefore an EMD era has a non-trivial impact on the values of the favoured reheating temperature and on the allowed DM particle mass. In this spirit, in this work we revisited and generalized the viable parameter space of the two simple EFT models studied in \cite{Goudelis:2022bls} considering the cosmological scenario of a transient EMD era.
 
Such a transient non-thermal era can also have a visible imprint on CMB observables because it modifies the cosmic expansion rate and, consequently, the rate at which modes $k$ re-enter the horizon and the CMB measured $n_s(k)$ value. The spectral index $n_s$ and tensor-to-scalar ratio values can be predicted in the framework of explicit inflationary models. Although the dependence of inflationary observables on the number of e-folds $N_*$ is certainly well-known, it is not very common to study explicit particle physics models in conjuction with inflation in such a context. 

Within our scenarios of simultaneous DM and baryon asymmetry production, the allowed values for the reheating temperature and the dilution size are correlated, given a specific microphysical framework. In light of this observation, we examined the implications of these scenarios for typical inflationary models. We computed the allowed values for the reheating temperature and the duration of the EMD era for the two cases of specific EFT models and showed that, once embedded within concrete inflationary scenarios, the CMB data can constrain these models and, therefore, provide a phenomenological handle on constructions which are, otherwise, extremely hard to test. Our analysis demonstrates at a quantitative level that such a unified description provides a promising, though challenging, strategy for model selection within complete scenarios of cosmic evolution.

As a final remark let us note that such a unified description can be further tested by a variety of remarkably different experimental probes: from  gravitational wave experiments that could detect gravitational radiation and probe a potential EMD era  \cite{Kuroyanagi:2014qza, Inomata:2019zqy, Dalianis:2020gup}, CMB-Stage 4 experiments \cite{ CMB-S4:2020lpa} that will have the sensitivity to detect a $r>0.003$, signal, and experiments such as EUCLID \cite{EuclidTheoryWorkingGroup:2012gxx} and cosmic 21-cm surveys \cite{Mao:2008ug, Pritchard:2011xb} that can achieve a precision of $10^{-3}$ in the value of $n_s$, to particle physics experiments that search for BSM physics and DM particles, see \textit{e.g.} \cite{Roszkowski:2017nbc, Belanger:2018sti}.

\section*{Acknowledgements}
 The work of I.D. and V.C.S. was supported by the Hellenic Foundation for Research and Innovation (H.F.R.I.) under the ``First Call for H.F.R.I. Research Projects to support Faculty members and Researchers and the procurement of high-cost research equipment grant'' (Project Number: 824). This research is co-financed by Greece and the European Union (European Social Fund- ESF) through the  Operational Programme ``Human Resources Development, Education and Lifelong Learning" in the context of the project ``Strengthening Human Resources Research Potential via Doctorate Research - 2nd Cycle" (MIS-5000432), implemented by the State Scholarships Foundation (IKY). The work of D.K. is supported by the Lancaster-Manchester-Sheffield Consortium for Fundamental Physics, under STFC Research Grant ST/T001038/1.

\bibliographystyle{JHEP}
\bibliography{bibliography}

\end{document}